\documentclass[10pt]{article}
\usepackage{graphicx}
\usepackage{amsmath}
\usepackage{amssymb}
\usepackage{caption2}
\begin{document}
\bibliographystyle{prsty}
\begin{center}
{\large {\bf \sc{  Analysis of  the $P_c(4312)$,  $P_c(4440)$, $P_c(4457)$ and related hidden-charm  pentaquark states  with QCD sum rules }}} \\[2mm]
Zhi-Gang Wang \footnote{E-mail: zgwang@aliyun.com.  }     \\
 Department of Physics, North China Electric Power University, Baoding 071003, P. R. China
\end{center}

\begin{abstract}
 In this article, we restudy the ground state mass spectrum of the diquark-diquark-antiquark type $uudc\bar{c}$ pentaquark states with the QCD sum rules by carrying out the operator product expansion up to the   vacuum condensates of $13$ in a consistent way. The predicted masses  support assigning the $P_c(4312)$ to be the hidden-charm pentaquark state with  $J^{P}={\frac{1}{2}}^-$, assigning the $P_c(4440)$ to be the hidden-charm pentaquark state with  $J^{P}={\frac{1}{2}}^-$, ${\frac{3}{2}}^-$ or ${\frac{5}{2}}^-$, assigning the $P_c(4457)$ to be the hidden-charm pentaquark state with $J^{P}={\frac{1}{2}}^-$ or ${\frac{3}{2}}^-$.
\end{abstract}

 PACS number: 12.39.Mk, 14.20.Lq, 12.38.Lg

Key words: Pentaquark states, QCD sum rules

\section{Introduction}
In 2015,  the  LHCb collaboration studied the $\Lambda_b^0\to J/\psi K^- p$ decays  and observed  two pentaquark candidates $P_c(4380)$ and $P_c(4450)$ in the $J/\psi p$ mass spectrum     with the significances of more than $9\sigma$  \cite{LHCb-4380}.
Recently, the LHCb collaboration studied the $\Lambda_b^0\to J/\psi K^- p$ decays with a data sample, which is an order of magnitude larger than that previously analyzed by the LHCb collaboration, and observed a  narrow pentaquark candidate $P_c(4312)$  with a statistical significance of  $7.3\sigma$ \cite{LHCb-Pc4312}. Furthermore,
 the LHCb collaboration confirmed the $P_c(4450)$ pentaquark structure, and observed that it consists  of two narrow overlapping peaks $P_c(4440)$ and $P_c(4457)$
  with  the statistical significance of  $5.4\sigma$ \cite{LHCb-Pc4312}.
   The measured  masses and widths are
\begin{flalign}
 &P_c(4312) : M = 4311.9\pm0.7^{+6.8}_{-0.6} \mbox{ MeV}\, , \, \Gamma = 9.8\pm2.7^{+ 3.7}_{- 4.5} \mbox{ MeV} \, , \nonumber \\
 & P_c(4440) : M = 4440.3\pm1.3^{+4.1}_{-4.7} \mbox{ MeV}\, , \, \Gamma = 20.6\pm4.9_{-10.1}^{+ 8.7} \mbox{ MeV} \, , \nonumber \\
 &P_c(4457) : M = 4457.3\pm0.6^{+4.1}_{-1.7} \mbox{ MeV} \, ,\, \Gamma = 6.4\pm2.0_{- 1.9}^{+ 5.7} \mbox{ MeV} \,   .
\end{flalign}
 There have been  several  possible  assignments of the $P_c$ states since the observations of the $P_c(4380)$ and $P_c(4450)$, such as the diquark-diquark-antiquark type pentaquark states \cite{di-di-anti-penta-1,di-di-anti-penta-2,di-di-anti-penta-3,di-di-anti-penta-4,di-di-anti-penta-5,Wang1508-EPJC,WangHuang-EPJC-1508-12,
 WangZG-EPJC-1509-12,WangZG-NPB-1512-32,WangZhang-APPB,Pc4312-penta-1,Pc4312-penta-2,Pc4312-penta-3},  the diquark-triquark  type  pentaquark states \cite{di-tri-penta-1, di-tri-penta-2,di-tri-penta-3},
 the  molecule-like  pentaquark  states \cite{mole-penta-1,mole-penta-2,mole-penta-3,mole-penta-4,mole-penta-5,mole-penta-6,mole-penta-7,mole-penta-8,mole-penta-9,mole-penta-10,WangPenta-IJMPA,
Pc4312-mole-penta-1,Pc4312-mole-penta-2,Pc4312-mole-penta-3,Pc4312-mole-penta-4,Pc4312-mole-penta-5,Pc4312-mole-penta-6,Pc4312-mole-penta-7,Pc4312-mole-penta-8,
Pc4312-mole-penta-9,Pc4312-mole-penta-10}, the hadro-charmonium states \cite{Pc4312-hadrocharmonium},
 the re-scattering effects \cite{rescattering-penta-1,rescattering-penta-2}, etc.
In this article, we choose the diquark-diquark-antiquark type pentaquark scenario, and restudy the ground state mass spectrum of the pentaquark states with the QCD sum rules.

The QCD sum rules is a powerful theoretical tool in studying the properties of the ground state mesons and  baryons, such as the masses, decay constants, form-factors, hadronic coupling constants \cite{SVZ79,PRT85}. In the  QCD sum rules, the operator product expansion is used to expand the
time-ordered currents into a series of quark and gluon condensates which parameterize the nonperturbative  properties of the QCD vacuum.
Based on the quark-hadron duality, we can obtain copious information
about the hadronic parameters at the phenomenological side
\cite{SVZ79,PRT85}.

There have been many works on the mass spectrum  of
the exotic states  $X$, $Y$, $Z$ and $P$ \cite{Nielsen-Review}.
 In Refs.\cite{Wang1508-EPJC,WangHuang-EPJC-1508-12,WangZG-EPJC-1509-12,WangZG-NPB-1512-32,WangZhang-APPB}, we construct the diquark-diquark-antiquark type pentaquark currents, study the $J^P={\frac{1}{2}}^\pm$, ${\frac{3}{2}}^\pm$, ${\frac{5}{2}}^\pm$   hidden-charm pentaquark states with the strangeness  $S=0,\,-1,\,-2,\,-3$ systematically using the QCD sum rules, and explore the possible assignment of the $P_c(4380)$ and $P_c(4450)$ in the  scenario of the pentaquark states.
 In carrying out the operator product  expansion, we take into account the contributions of the vacuum condensates which are vacuum expectations of the quark-gluon operators of the order $\mathcal{O}(\alpha_s^k)$ with $k\leq1$ and dimension $D\leq 10$, and use the energy scale formula $\mu=\sqrt{M_{P}^2-(2{\mathbb{M}}_c)^2}$  with the old value ${\mathbb{M}}_c=1.80\,\rm{GeV}$ of the effective $c$-quark mass to determine the ideal energy scales of the QCD spectral densities.

  In this article, we restudy the ground state mass spectrum of the diquark-diquark-antiquark type $uudc\bar{c}$ pentaquark states with the QCD sum rules by taking into account  all the  vacuum condensates up to the quark-gluon operators of the order $\mathcal{O}(\alpha_s^k)$ with $k\leq1$  and dimension $13$  in  carrying out the operator product expansion, and use the energy scale formula $\mu=\sqrt{M_{P}^2-(2{\mathbb{M}}_c)^2}$  with the updated value ${\mathbb{M}}_c=1.82\,\rm{GeV}$ to determine the ideal energy scales of the QCD spectral densities  \cite{WangEPJC-1601}, and update the analysis and explore the possible assignments of the $P_c(4312)$, $P_c(4440)$ and $P_c(4457)$ in the scenario of the pentaquark states.

In Ref.\cite{WangPenta-IJMPA}, we choose the color-singlet-color-singlet type or meson-baryon type  currents to interpolate the $\bar{D}\Sigma$,
 $\bar{D}\Sigma^*$, $\bar{D}^*\Sigma$ and $\bar{D}^*\Sigma^*$ pentaquark molecular states, and observe that the experimental values of the masses of the LHCb pentaquark candidates $P_c(4312)$,  $P_c(4440)$ and  $P_c(4457)$ can be reproduced  in the meson-baryon molecule scenario. In this article, we explore  the relation between the (compact) pentaquark scenario and molecule scenario.

 The article is arranged as follows:
  we derive the QCD sum rules for the masses and pole residues of  the
ground state hidden-charm pentaquark states in Sect.2;  in Sect.3, we present the numerical results and discussions; and Sect.4 is reserved for our
conclusion.

\section{QCD sum rules for  the  hidden-charm pentaquark states}
Firstly, we write down  the two-point correlation functions $\Pi(p)$, $\Pi_{\mu\nu}(p)$ and $\Pi_{\mu\nu\alpha\beta}(p)$  in the QCD sum rules,
\begin{eqnarray}\label{CF-Pi-Pi-Pi}
\Pi(p)&=&i\int d^4x e^{ip \cdot x} \langle0|T\left\{J(x)\bar{J}(0)\right\}|0\rangle \, , \nonumber \\
\Pi_{\mu\nu}(p)&=&i\int d^4x e^{ip \cdot x} \langle0|T\left\{J_{\mu}(x)\bar{J}_{\nu}(0)\right\}|0\rangle \, , \nonumber \\
\Pi_{\mu\nu\alpha\beta}(p)&=&i\int d^4x e^{ip \cdot x} \langle0|T\left\{J_{\mu\nu}(x)\bar{J}_{\alpha\beta}(0)\right\}|0\rangle \, ,
\end{eqnarray}
where the currents $J(x)=J^1(x)$, $J^2(x)$, $J^3(x)$, $J^4(x)$,  $J_\mu(x)=J^1_\mu(x)$, $J^2_\mu(x)$, $J^3_\mu(x)$, $J^4_\mu(x)$, $J_{\mu\nu}(x)=J^1_{\mu\nu}(x)$, $J^2_{\mu\nu}(x)$,
\begin{eqnarray}
 J^1(x)&=&\varepsilon^{ila} \varepsilon^{ijk}\varepsilon^{lmn}  u^T_j(x) C\gamma_5 d_k(x)\,u^T_m(x) C\gamma_5 c_n(x)\,  C\bar{c}^{T}_{a}(x) \, , \nonumber\\
J^2(x)&=&\varepsilon^{ila} \varepsilon^{ijk}\varepsilon^{lmn}  u^T_j(x) C\gamma_5 d_k(x)\,u^T_m(x) C\gamma_\mu c_n(x)\,\gamma_5 \gamma^\mu C\bar{c}^{T}_{a}(x) \, , \nonumber\\
J^{3}(x)&=&\frac{\varepsilon^{ila} \varepsilon^{ijk}\varepsilon^{lmn}}{\sqrt{3}} \left[ u^T_j(x) C\gamma_\mu u_k(x) d^T_m(x) C\gamma_5 c_n(x)+2u^T_j(x) C\gamma_\mu d_k(x) u^T_m(x) C\gamma_5 c_n(x)\right] \gamma_5 \gamma^\mu  C\bar{c}^{T}_{a}(x) \, ,  \nonumber\\
J^{4}(x)&=&\frac{\varepsilon^{ila} \varepsilon^{ijk}\varepsilon^{lmn}}{\sqrt{3}} \left[ u^T_j(x) C\gamma_\mu u_k(x)d^T_m(x) C\gamma^\mu c_n(x)+2u^T_j(x) C\gamma_\mu d_k(x)u^T_m(x) C\gamma^\mu c_n(x) \right] C\bar{c}^{T}_{a}(x) \, , \nonumber\\
J^1_{\mu}(x)&=&\varepsilon^{ila} \varepsilon^{ijk}\varepsilon^{lmn}  u^T_j(x) C\gamma_5 d_k(x)\,u^T_m(x) C\gamma_\mu c_n(x)\, C\bar{c}^{T}_{a}(x) \, , \nonumber \\
J^{2}_{\mu}(x)&=&\frac{\varepsilon^{ila} \varepsilon^{ijk}\varepsilon^{lmn}}{\sqrt{3}} \left[ u^T_j(x) C\gamma_\mu u_k(x) d^T_m(x) C\gamma_5 c_n(x)+2u^T_j(x) C\gamma_\mu d_k(x) u^T_m(x) C\gamma_5 c_n(x)\right]    C\bar{c}^{T}_{a}(x) \, ,  \nonumber\\
J^{3}_{\mu}(x)&=&\frac{\varepsilon^{ila} \varepsilon^{ijk}\varepsilon^{lmn}}{\sqrt{3}} \left[ u^T_j(x) C\gamma_\mu u_k(x)d^T_m(x) C\gamma_\alpha c_n(x)+2u^T_j(x) C\gamma_\mu d_k(x)u^T_m(x) C\gamma_\alpha c_n(x) \right] \gamma_5\gamma^\alpha C\bar{c}^{T}_{a}(x) \, , \nonumber\\
 J^{4}_{\mu}(x)&=&\frac{\varepsilon^{ila} \varepsilon^{ijk}\varepsilon^{lmn}}{\sqrt{3}} \left[ u^T_j(x) C\gamma_\alpha u_k(x)d^T_m(x) C\gamma_\mu c_n(x)+2u^T_j(x) C\gamma_\alpha d_k(x)u^T_m(x) C\gamma_\mu c_n(x) \right] \gamma_5\gamma^\alpha C\bar{c}^{T}_{a}(x) \, , \nonumber
 \end{eqnarray}
 \begin{eqnarray}
J^1_{\mu\nu}(x)&=&\frac{\varepsilon^{ila} \varepsilon^{ijk}\varepsilon^{lmn}}{\sqrt{6}} \left[ u^T_j(x) C\gamma_\mu u_k(x)d^T_m(x) C\gamma_\nu c_n(x)+2u^T_j(x) C\gamma_\mu d_k(x)u^T_m(x) C\gamma_\nu c_n(x) \right]    \nonumber\\
 &&C\bar{c}^{T}_{a}(x)+\left( \mu\leftrightarrow\nu\right)\, ,  \nonumber\\
J^2_{\mu\nu}(x)&=&\frac{1}{\sqrt{2}}\varepsilon^{ila} \varepsilon^{ijk}\varepsilon^{lmn}  u^T_j(x) C\gamma_5 d_k(x)\left[u^T_m(x) C\gamma_\mu c_n(x)\, \gamma_5\gamma_{\nu}C\bar{c}^{T}_{a}(x)\right.\nonumber\\
&&\left.+u^T_m(x) C\gamma_\nu c_n(x)\,\gamma_5 \gamma_{\mu}C\bar{c}^{T}_{a}(x)\right] \, ,
\end{eqnarray}
where the $i$, $j$, $k$, $l$, $m$, $n$ and $a$ are color indices, the $C$ is the charge conjugation matrix \cite{Wang1508-EPJC,WangHuang-EPJC-1508-12,WangZG-EPJC-1509-12,WangZG-NPB-1512-32}. The attractive interaction induced by one-gluon exchange favors  formation of diquark correlations in color antitriplet $\bar{3}_c$ channels, we prefer  the diquark operators in color antitriplet $\bar{3}_c$.
Compared to the pseudoscalar and vector diquark states, the scalar and axialvector diquark states are  favored configurations,  we choose the scalar and axialvector diquark  operators in color antitriplet $\bar{3}_c$  as the basic constituents to construct the  diquark-diquark-antiquark type current operators $J(x)$, $J_\mu(x)$ and $J_{\mu\nu}(x)$ with the spin-parity $J^{P}={\frac{1}{2}}^-$, ${\frac{3}{2}}^-$ and ${\frac{5}{2}}^-$, respectively,  which are expected to couple potentially to the lowest pentaquark states \cite{Wang1508-EPJC,WangHuang-EPJC-1508-12,WangZG-EPJC-1509-12,WangZG-NPB-1512-32}.

In the currents $J(x)$, $J_\mu(x)$ and $J_{\mu\nu}(x)$, there are diquark operators $\varepsilon^{ijk}u^T_jC\gamma_5d_k$, $\varepsilon^{ijk}u^T_jC\gamma_{\mu}d_k$, $\varepsilon^{ijk}u^T_jC\gamma_{\mu}u_k$, $\varepsilon^{ijk}q^T_jC\gamma_5c_k$, $\varepsilon^{ijk}q^T_jC\gamma_{\mu}c_k$ with $q=u$, $d$. If we use the $S_L$ and $S_H$
to denote the spins of the light diquarks and heavy diquarks respectively, the light diquark operators  $\varepsilon^{ijk}u^T_jC\gamma_5d_k$, $\varepsilon^{ijk}u^T_jC\gamma_{\mu}d_k$ and $\varepsilon^{ijk}u^T_jC\gamma_{\mu}u_k$ have the spins $S_L=0$, $1$ and $1$, respectively, while the heavy diquark operators $\varepsilon^{ijk}q^T_jC\gamma_5c_k$ and $\varepsilon^{ijk}q^T_jC\gamma_{\mu}c_k$ have the spins $S_H=0$ and $1$, respectively. The light diquark and  heavy diquark form a charmed tetraquark in color triplet with the  angular momentum $\vec{J}_{LH}=\vec{S}_L+\vec{S}_H$, which has the values $J_{LH}=0$, $1$ or $2$.
The $\bar{c}$-quark operator $Cc_a^T$ has the spin-parity $J^P={\frac{1}{2}}^-$,
while the $\bar{c}$-quark operator $\gamma_5\gamma_{\mu}Cc_a^T$ has the spin-parity $J^P={\frac{3}{2}}^-$ due to  the axialvector-like factor $\gamma_5\gamma_{\mu}$. The total angular momentums  of the currents are $\vec{J}=\vec{J}_{LH}+\vec{J}_{\bar{c}}$ with the values $J=\frac{1}{2}$, $\frac{3}{2}$ or $\frac{5}{2}$, which are shown explicitly in Table \ref{current-pentaQ}. In Table \ref{current-pentaQ}, we present the quark structures of the interpolating  currents   explicitly. For example, in the current operator $J^2_{\mu\nu}(x)$, there are a scalar diquark operator
$\varepsilon^{ijk}  u^T_j(x) C\gamma_5 d_k(x)$ with the spin-parity $J^P=0^+$, an axialvector diquark operator $\varepsilon^{lmn} u^T_m(x) C\gamma_\mu c_n(x)$ with the spin-parity $J^P=1^+$, and an antiquark operator $\gamma_5\gamma_{\nu}C\bar{c}^T_a(x)$ with the spin-parity $J^{P}={\frac{3}{2}}^-$,  the total angular momentum of the current is $J={\frac{5}{2}}$.

\begin{table}
\begin{center}
\begin{tabular}{|c|c|c|c|c|c|c|c|c|}\hline\hline
$[qq^\prime][q^{\prime\prime}c]\bar{c}$ ($S_L$, $S_H$, $J_{LH}$, $J$)& $J^{P}$              & Currents              \\ \hline

$[ud][uc]\bar{c}$ ($0$, $0$, $0$, $\frac{1}{2}$)                     & ${\frac{1}{2}}^{-}$  & $J^1(x)$              \\

$[ud][uc]\bar{c}$ ($0$, $1$, $1$, $\frac{1}{2}$)                     & ${\frac{1}{2}}^{-}$  & $J^2(x)$              \\

$[uu][dc]\bar{c}+2[ud][uc]\bar{c}$ ($1$, $0$, $1$, $\frac{1}{2}$)    & ${\frac{1}{2}}^{-}$  & $J^3(x)$              \\

$[uu][dc]\bar{c}+2[ud][uc]\bar{c}$ ($1$, $1$, $0$, $\frac{1}{2}$)    & ${\frac{1}{2}}^{-}$  & $J^4(x)$              \\

$[ud][uc]\bar{c}$ ($0$, $1$, $1$, $\frac{3}{2}$)                     & ${\frac{3}{2}}^{-}$  & $J^1_\mu(x)$           \\

$[uu][dc]\bar{c}+2[ud][uc]\bar{c}$ ($1$, $0$, $1$, $\frac{3}{2}$)    & ${\frac{3}{2}}^{-}$  & $J^2_\mu(x)$          \\

$[uu][dc]\bar{c}+2[ud][uc]\bar{c}$ ($1$, $1$, $2$, $\frac{3}{2}$)    & ${\frac{3}{2}}^{-}$  & $J^3_\mu(x)$           \\

$[uu][dc]\bar{c}+2[ud][uc]\bar{c}$ ($1$, $1$, $2$, $\frac{3}{2}$)    & ${\frac{3}{2}}^{-}$  & $J^4_\mu(x)$           \\

$[uu][dc]\bar{c}+2[ud][uc]\bar{c}$ ($1$, $1$, $2$, $\frac{5}{2}$)    & ${\frac{5}{2}}^{-}$  & $J^1_{\mu\nu}(x)$           \\

$[ud][uc]\bar{c}$ ($0$, $1$, $1$, $\frac{5}{2}$)                     & ${\frac{5}{2}}^{-}$  & $J^2_{\mu\nu}(x)$     \\ \hline\hline
\end{tabular}
\end{center}
\caption{ The quark structures of the  current operators, where the $S_L$ and $S_H$ denote the spins of the light diquarks and heavy diquarks respectively, $\vec{J}_{LH}=\vec{S}_L+\vec{S}_H$, $\vec{J}=\vec{J}_{LH}+\vec{J}_{\bar{c}}$, the $\vec{J}_{\bar{c}}$ is the angular momentum of the $\bar{c}$-quark.
As the current operators couple potentially to pentaquark states which have the same quark structures, thereafter we will use the quark structures of the current operators to represent the corresponding pentaquark states.  }\label{current-pentaQ}
\end{table}

Although the currents $J(x)$, $J_\mu(x)$ and $J_{\mu\nu}(x)$ have negative parity, they also couple potentially to  the  positive parity pentaquark states, as   multiplying $i \gamma_{5}$ to the currents  $J(x)$, $J_\mu(x)$ and $J_{\mu\nu}(x)$ changes their parity \cite{Chung82,Bagan93,Oka96,WangHbaryon-1,WangHbaryon-2,WangHbaryon-3,WangHbaryon-4,WangHbaryon-5}.

Now we write down the current-pentaquark couplings (or the definitions for the pole residues) explicitly,
\begin{eqnarray}\label{Coupling12}
\langle 0| J (0)|P_{\frac{1}{2}}^{-}(p)\rangle &=&\lambda^{-}_{\frac{1}{2}} U^{-}(p,s) \, , \nonumber \\
\langle 0| J (0)|P_{\frac{1}{2}}^{+}(p)\rangle &=&\lambda^{+}_{\frac{1}{2}} i\gamma_5 U^{+}(p,s) \, ,
\end{eqnarray}
\begin{eqnarray}
\langle 0| J_{\mu} (0)|P_{\frac{3}{2}}^{-}(p)\rangle &=&\lambda^{-}_{\frac{3}{2}} U^{-}_\mu(p,s) \, ,  \nonumber \\
\langle 0| J_{\mu} (0)|P_{\frac{3}{2}}^{+}(p)\rangle &=&\lambda^{+}_{\frac{3}{2}}i\gamma_5 U^{+}_\mu(p,s) \, ,  \nonumber \\
\langle 0| J_{\mu} (0)|P_{\frac{1}{2}}^{+}(p)\rangle &=&f^{+}_{\frac{1}{2}}p_\mu U^{+}(p,s) \, , \nonumber \\
\langle 0| J_{\mu} (0)|P_{\frac{1}{2}}^{-}(p)\rangle &=&f^{-}_{\frac{1}{2}}p_\mu i\gamma_5 U^{-}(p,s) \, ,
\end{eqnarray}
\begin{eqnarray}\label{Coupling52}
\langle 0| J_{\mu\nu} (0)|P_{\frac{5}{2}}^{-}(p)\rangle &=&\sqrt{2}\lambda^{-}_{\frac{5}{2}} U^{-}_{\mu\nu}(p,s) \, ,\nonumber\\
\langle 0| J_{\mu\nu} (0)|P_{\frac{5}{2}}^{+}(p)\rangle &=&\sqrt{2}\lambda^{+}_{\frac{5}{2}}i\gamma_5 U^{+}_{\mu\nu}(p,s) \, ,\nonumber\\
\langle 0| J_{\mu\nu} (0)|P_{\frac{3}{2}}^{+}(p)\rangle &=&f^{+}_{\frac{3}{2}} \left[p_\mu U^{+}_{\nu}(p,s)+p_\nu U^{+}_{\mu}(p,s)\right] \, , \nonumber\\
\langle 0| J_{\mu\nu} (0)|P_{\frac{3}{2}}^{-}(p)\rangle &=&f^{-}_{\frac{3}{2}}i\gamma_5 \left[p_\mu U^{-}_{\nu}(p,s)+p_\nu U^{-}_{\mu}(p,s)\right] \, , \nonumber\\
\langle 0| J_{\mu\nu} (0)|P_{\frac{1}{2}}^{-}(p)\rangle &=&g^{-}_{\frac{1}{2}}p_\mu p_\nu U^{-}(p,s) \, , \nonumber\\
\langle 0| J_{\mu\nu} (0)|P_{\frac{1}{2}}^{+}(p)\rangle &=&g^{+}_{\frac{1}{2}}p_\mu p_\nu i\gamma_5 U^{+}(p,s) \, ,
\end{eqnarray}
where the superscripts $\pm$ denote the positive parity and negative parity, respectively, the subscripts $\frac{1}{2}$, $\frac{3}{2}$ and $\frac{5}{2}$ denote the spins of the pentaquark states,     the $\lambda$, $f$ and $g$ are the pole residues.
The spinors $U^\pm(p,s)$ satisfy the Dirac equations  $(\not\!\!p-M_{\pm})U^{\pm}(p)=0$, while the spinors $U^{\pm}_\mu(p,s)$ and $U^{\pm}_{\mu\nu}(p,s)$ satisfy the Rarita-Schwinger equations $(\not\!\!p-M_{\pm})U^{\pm}_\mu(p)=0$ and $(\not\!\!p-M_{\pm})U^{\pm}_{\mu\nu}(p)=0$,  and the relations $\gamma^\mu U^{\pm}_\mu(p,s)=0$,
$p^\mu U^{\pm}_\mu(p,s)=0$, $\gamma^\mu U^{\pm}_{\mu\nu}(p,s)=0$,
$p^\mu U^{\pm}_{\mu\nu}(p,s)=0$, $ U^{\pm}_{\mu\nu}(p,s)= U^{\pm}_{\nu\mu}(p,s)$, respectively. For more details about the spinors, one can consult Ref.\cite{Wang1508-EPJC}.

 At the phenomenological  side, we insert  a complete set  of intermediate pentaquark states with the same quantum numbers as the current operators  $J(x)$, $i\gamma_5 J(x)$, $J_\mu(x)$, $i\gamma_5 J_\mu(x)$, $J_{\mu\nu}(x)$ and $i\gamma_5 J_{\mu\nu}(x)$ into the correlation functions
$\Pi(p)$, $\Pi_{\mu\nu}(p)$ and $\Pi_{\mu\nu\alpha\beta}(p)$ to obtain the hadronic representation
\cite{SVZ79,PRT85}. We take into account the current-pentaquark couplings (or the quark-hadron duality) shown in Eqs.\eqref{Coupling12}-\eqref{Coupling52}, and  isolate the pole terms of the lowest
states of the negative parity and positive parity hidden-charm  pentaquark states, and obtain the
following results:
\begin{eqnarray}\label{CF-Hadron-12}
\Pi(p) & = & {\lambda^{-}_{\frac{1}{2}}}^2  {\!\not\!{p}+ M_{-} \over M_{-}^{2}-p^{2}  }+  {\lambda^{+}_{\frac{1}{2}}}^2  {\!\not\!{p}- M_{+} \over M_{+}^{2}-p^{2}  } +\cdots  \, ,\nonumber\\
&=&\Pi_{\frac{1}{2}}^1(p^2)\!\not\!{p}+\Pi_{\frac{1}{2}}^0(p^2)\, ,
 \end{eqnarray}
\begin{eqnarray}\label{CF-Hadron-32}
\Pi_{\mu\nu}(p) & = & \left({\lambda^{-}_{\frac{3}{2}}}^2  {\!\not\!{p}+ M_{-} \over M_{-}^{2}-p^{2}  }+ {\lambda^{+}_{\frac{3}{2}}}^2  {\!\not\!{p}- M_{+} \over M_{+}^{2}-p^{2}  }\right) \left(- g_{\mu\nu}\right)+\cdots  \, ,\nonumber\\
&=&\left[\Pi_{\frac{3}{2}}^1(p^2)\!\not\!{p}+\Pi_{\frac{3}{2}}^0(p^2)\right]\left(- g_{\mu\nu}\right)+\cdots\, ,
\end{eqnarray}
\begin{eqnarray}\label{CF-Hadron-52}
\Pi_{\mu\nu\alpha\beta}(p) & = & \left({\lambda^{-}_{\frac{5}{2}}}^2  {\!\not\!{p}+ M_{-} \over M_{-}^{2}-p^{2}  } +{\lambda^{+}_{\frac{5}{2}}}^2  {\!\not\!{p}- M_{+} \over M_{+}^{2}-p^{2}  }\right)\left( g_{\mu\alpha}g_{\nu\beta}+g_{\mu\beta}g_{\nu\alpha}\right)  +\cdots \, , \nonumber\\
& = & \left[\Pi_{\frac{5}{2}}^1(p^2)\!\not\!{p}+\Pi_{\frac{5}{2}}^0(p^2)\right]\left( g_{\mu\alpha}g_{\nu\beta}+g_{\mu\beta}g_{\nu\alpha}\right)  +\cdots \, .
 \end{eqnarray}
In this article, we study the components $\Pi_{\frac{1}{2}}^1(p^2)$, $\Pi_{\frac{1}{2}}^0(p^2)$, $\Pi_{\frac{3}{2}}^1(p^2)$, $\Pi_{\frac{3}{2}}^0(p^2)$, $\Pi_{\frac{5}{2}}^1(p^2)$, $\Pi_{\frac{5}{2}}^0(p^2)$ to avoid possible contaminations from other pentaquark states with different spins. For detailed discussions about this subject, one can consult Refs.\cite{Wang1508-EPJC,Wang-cc-baryon-penta}.

Now we take a digression  to discuss the relation between the (compact) pentaquark scenario and molecule scenario.
 In this article, we study the mass spectrum of the diquark-diquark-antiquark type pentaquark states with the QCD sum rules. The diquark-diquark-antiquark type pentaquark current operator  with special quantum numbers couples potentially  to a special pentaquark state,
 while the current operator  can be re-arranged both in the color and Dirac-spinor  spaces, and changed  to a current operator as a special superposition of
 a series of color-singlet-color-singlet type (baryon-meson type) current operators.
  We perform Fierz rearrangements for  the currents $J(x)$, $J_\mu(x)$ and $J_{\mu\nu}(x)$   to   obtain the results,
\begin{eqnarray} \label{Fierz-J1}
J^1&=&-\frac{1}{4}\mathcal{S}_{ud}\gamma_5c\,\bar{c}u+\frac{1}{4}\mathcal{S}_{ud}\gamma^\lambda\gamma_5c\,\bar{c}\gamma_{\lambda}u
+\frac{1}{8}\mathcal{S}_{ud}\sigma^{\lambda\tau}\gamma_5c\,\bar{c}\sigma_{\lambda\tau}u+\frac{1}{4}\mathcal{S}_{ud}\gamma^{\lambda}c\,\bar{c}\gamma_{\lambda}\gamma_5u
+\frac{i}{4}\mathcal{S}_{ud}c\,\bar{c}i\gamma_5 u \nonumber\\
&&+\frac{1}{4}\mathcal{S}_{ud}\gamma_5u\,\bar{c}c-\frac{1}{4}\mathcal{S}_{ud}\gamma^\lambda\gamma_5u\,\bar{c}\gamma_{\lambda}c
-\frac{1}{8}\mathcal{S}_{ud}\sigma^{\lambda\tau}\gamma_5u\,\bar{c}\sigma_{\lambda\tau}c-\frac{1}{4}\mathcal{S}_{ud}\gamma^{\lambda}u\,\bar{c}\gamma_{\lambda}\gamma_5c
-\frac{i}{4}\mathcal{S}_{ud}u\,\bar{c}i\gamma_5 c\, , \nonumber\\
\end{eqnarray}

\begin{eqnarray}
J^2&=&-\mathcal{S}_{ud}\gamma_5c\,\bar{c}u+\frac{1}{2}\mathcal{S}_{ud}\gamma^\lambda\gamma_5c\,\bar{c}\gamma_{\lambda}u
-\frac{1}{2}\mathcal{S}_{ud}\gamma^{\lambda}c\,\bar{c}\gamma_{\lambda}\gamma_5u-i\mathcal{S}_{ud}c\,\bar{c}i\gamma_5 u -\mathcal{S}_{ud}\gamma_5u\,\bar{c}c\nonumber\\
&&+\frac{1}{2}\mathcal{S}_{ud}\gamma^\lambda\gamma_5u\,\bar{c}\gamma_{\lambda}c
-\frac{1}{2}\mathcal{S}_{ud}\gamma^{\lambda}u\,\bar{c}\gamma_{\lambda}\gamma_5c
-i\mathcal{S}_{ud}u\,\bar{c}i\gamma_5 c\, ,
\end{eqnarray}

\begin{eqnarray}
\sqrt{3}J^3&=&\frac{1}{4}\mathcal{S}^\mu_{uu}\gamma_{\mu}c\,\bar{c}d+\frac{1}{4}\mathcal{S}^\mu_{uu}\gamma_{\mu}\gamma_{\lambda}c\,\bar{c}\gamma^{\lambda}d
-\frac{1}{8}\mathcal{S}^\mu_{uu}\gamma_{\mu}\sigma_{\lambda\tau}c\,\bar{c}\sigma^{\lambda\tau}d
+\frac{1}{4}\mathcal{S}^\mu_{uu}\gamma_{\mu}\gamma_{\lambda}\gamma_5c\,\bar{c}\gamma^{\lambda}\gamma_5d\nonumber\\
&&-\frac{i}{4}\mathcal{S}^\mu_{uu}\gamma_{\mu}\gamma_{5}c\,\bar{c}i\gamma_{5}d
-\frac{1}{4}\mathcal{S}^\mu_{uu}\gamma_{\mu}d\,\bar{c}c-\frac{1}{4}\mathcal{S}^\mu_{uu}\gamma_{\mu}\gamma_{\lambda}d\,\bar{c}\gamma^{\lambda}c
+\frac{1}{8}\mathcal{S}^\mu_{uu}\gamma_{\mu}\sigma_{\lambda\tau}d\,\bar{c}\sigma^{\lambda\tau}c\nonumber\\
&&-\frac{1}{4}\mathcal{S}^\mu_{uu}\gamma_{\mu}\gamma_{\lambda}\gamma_5d\,\bar{c}\gamma^{\lambda}\gamma_5c
+\frac{i}{4}\mathcal{S}^\mu_{uu}\gamma_{\mu}\gamma_{5}d\,\bar{c}i\gamma_{5}c+2\left(S^\mu_{uu} \to S^\mu_{ud},\, d \to u \right)\, ,
\end{eqnarray}

\begin{eqnarray}
\sqrt{3}J^4&=&-\frac{1}{4}\mathcal{S}^\mu_{uu}\gamma_{\mu}c\,\bar{c}d+\frac{1}{4}\mathcal{S}^\mu_{uu}\gamma_{\lambda}\gamma_{\mu}c\,\bar{c}\gamma^{\lambda}d
+\frac{1}{8}\mathcal{S}^\mu_{uu}\sigma_{\lambda\tau}\gamma_{\mu}c\,\bar{c}\sigma^{\lambda\tau}d
-\frac{1}{4}\mathcal{S}^\mu_{uu}\gamma_{\lambda}\gamma_{\mu}\gamma_5c\,\bar{c}\gamma^{\lambda}\gamma_5d\nonumber\\
&&-\frac{i}{4}\mathcal{S}^\mu_{uu}\gamma_{\mu}\gamma_{5}c\,\bar{c}i\gamma_{5}d
-\frac{1}{4}\mathcal{S}^\mu_{uu}\gamma_{\mu}d\,\bar{c}c+\frac{1}{4}\mathcal{S}^\mu_{uu}\gamma_{\lambda}\gamma_{\mu}d\,\bar{c}\gamma^{\lambda}c
+\frac{1}{8}\mathcal{S}^\mu_{uu}\sigma_{\lambda\tau}\gamma_{\mu}d\,\bar{c}\sigma^{\lambda\tau}c\nonumber\\
&&-\frac{1}{4}\mathcal{S}^\mu_{uu}\gamma_{\lambda}\gamma_{\mu}\gamma_5d\,\bar{c}\gamma^{\lambda}\gamma_5c
-\frac{i}{4}\mathcal{S}^\mu_{uu}\gamma_{\mu}\gamma_{5}d\,\bar{c}i\gamma_{5}c+2\left(S^\mu_{uu} \to S^\mu_{ud},\, d \to u \right)\, ,
\end{eqnarray}

\begin{eqnarray}
J^1_\mu&=&-\frac{1}{4}\mathcal{S}_{ud} \gamma_{\mu}c\,\bar{c}u+\frac{1}{4}\mathcal{S}_{ud}\gamma_{\lambda}\gamma_{\mu}c\,\bar{c}\gamma^{\lambda}u
+\frac{1}{8}\mathcal{S}_{ud} \sigma_{\lambda\tau}\gamma_{\mu}c\,\bar{c}\sigma^{\lambda\tau}u
-\frac{1}{4}\mathcal{S}_{ud} \gamma_{\lambda}\gamma_{\mu}\gamma_5c\,\bar{c}\gamma^{\lambda}\gamma_5u\nonumber\\
&&-\frac{i}{4}\mathcal{S}_{ud} \gamma_{\mu}\gamma_{5}c\,\bar{c}i\gamma_{5}u
-\frac{1}{4}\mathcal{S}_{ud}\gamma_{\mu}u\,\bar{c}c+\frac{1}{4}\mathcal{S}_{ud}\gamma_{\lambda}\gamma_{\mu}u\,\bar{c}\gamma^{\lambda}c
+\frac{1}{8}\mathcal{S}_{ud}\sigma_{\lambda\tau}\gamma_{\mu}u\,\bar{c}\sigma^{\lambda\tau}c\nonumber\\
&&-\frac{1}{4}\mathcal{S}_{ud}\gamma_{\lambda}\gamma_{\mu}\gamma_5u\,\bar{c}\gamma^{\lambda}\gamma_5c
-\frac{i}{4}\mathcal{S}_{ud}\gamma_{\mu}\gamma_{5}u\,\bar{c}i\gamma_{5}c\, ,
\end{eqnarray}

\begin{eqnarray}
\sqrt{3}J^2_\mu&=&-\frac{1}{4}\mathcal{S}^{uu}_{\mu}\gamma_5c\,\bar{c}d+\frac{1}{4}\mathcal{S}^{uu}_{\mu}\gamma^\lambda\gamma_5c\,\bar{c}\gamma_{\lambda}d
+\frac{1}{8}\mathcal{S}^{uu}_{\mu}\sigma^{\lambda\tau}\gamma_5c\,\bar{c}\sigma_{\lambda\tau}d
+\frac{1}{4}\mathcal{S}^{uu}_{\mu}\gamma^{\lambda}c\,\bar{c}\gamma_{\lambda}\gamma_5d
+\frac{i}{4}\mathcal{S}^{uu}_{\mu}c\,\bar{c}i\gamma_5 d \nonumber\\
&&+\frac{1}{4}\mathcal{S}^{uu}_{\mu}\gamma_5d\,\bar{c}c-\frac{1}{4}\mathcal{S}^{uu}_{\mu}\gamma^\lambda\gamma_5d\,\bar{c}\gamma_{\lambda}c
-\frac{1}{8}\mathcal{S}^{uu}_{\mu}\sigma^{\lambda\tau}\gamma_5d\,\bar{c}\sigma_{\lambda\tau}c
-\frac{1}{4}\mathcal{S}^{uu}_{\mu}\gamma^{\lambda}d\,\bar{c}\gamma_{\lambda}\gamma_5c
-\frac{i}{4}\mathcal{S}^{uu}_{\mu}d\,\bar{c}i\gamma_5 c  \nonumber\\
&&+2\left(S_\mu^{uu} \to S_\mu^{ud},\, d \to u \right)\, ,
\end{eqnarray}

\begin{eqnarray}
\sqrt{3}J^3_\mu&=&-\mathcal{S}^{uu}_{\mu}\gamma_5c\,\bar{c}d
+\frac{1}{2}\mathcal{S}^{uu}_{\mu}\gamma^\lambda\gamma_5c\,\bar{c}\gamma_{\lambda}d
-\frac{1}{2}\mathcal{S}^{uu}_{\mu}\gamma^{\lambda}c\,\bar{c}\gamma_{\lambda}\gamma_5d
-i\mathcal{S}^{uu}_{\mu}c\,\bar{c}i\gamma_5 d-\mathcal{S}^{uu}_{\mu}\gamma_5d\,\bar{c}c \nonumber\\
&&+\frac{1}{2}\mathcal{S}^{uu}_{\mu}\gamma^\lambda\gamma_5d\,\bar{c}\gamma_{\lambda}c
-\frac{1}{2}\mathcal{S}^{uu}_{\mu}\gamma^{\lambda}d\,\bar{c}\gamma_{\lambda}\gamma_5c
-i\mathcal{S}^{uu}_{\mu}d\,\bar{c}i\gamma_5 c +2\left(S_\mu^{uu} \to S_\mu^{ud},\, d \to u \right)\, ,
\end{eqnarray}

\begin{eqnarray}
\sqrt{3}J^4_\mu&=&-\frac{1}{4}\mathcal{S}_{uu}^{\alpha}\gamma_{5}\gamma_{\alpha}\gamma_{\mu}c\,\bar{c}d
+\frac{1}{4}\mathcal{S}_{uu}^{\alpha}\gamma_{5}\gamma_{\alpha}c\,\bar{c}\gamma_{\mu}d
-\frac{i}{4}\mathcal{S}_{uu}^{\alpha}\gamma_{5}\gamma_{\alpha}\sigma_{\lambda\mu}c\,\bar{c}\gamma^{\lambda}d
+\frac{1}{8}\mathcal{S}_{uu}^{\alpha}\gamma_{5}\gamma_{\alpha}\sigma_{\lambda\tau}\gamma_{\mu}c\,\bar{c}\sigma^{\lambda\tau}d\nonumber\\
&&+\frac{1}{4}\mathcal{S}_{uu}^{\alpha}\gamma_{\alpha}c\,\bar{c}\gamma_{\mu}\gamma_{5}d
-\frac{i}{4}\mathcal{S}_{uu}^{\alpha}\gamma_{\alpha}\sigma_{\lambda\mu}c\,\bar{c}\gamma^{\lambda}\gamma_{5}d
-\frac{i}{4}\mathcal{S}_{uu}^{\alpha}\gamma_{\alpha}\gamma_{\mu}c\,\bar{c}i\gamma_{5}d
-\frac{1}{4}\mathcal{S}_{uu}^{\alpha}\gamma_{5}\gamma_{\alpha}\gamma_{\mu}d\,\bar{c}c\nonumber\\
&&+\frac{1}{4}\mathcal{S}_{uu}^{\alpha}\gamma_{5}\gamma_{\alpha}d\,\bar{c}\gamma_{\mu}c
-\frac{i}{4}\mathcal{S}_{uu}^{\alpha}\gamma_{5}\gamma_{\alpha}\sigma_{\lambda\mu}d\,\bar{c}\gamma^{\lambda}c
+\frac{1}{8}\mathcal{S}_{uu}^{\alpha}\gamma_{5}\gamma_{\alpha}\sigma_{\lambda\tau}\gamma_{\mu}d\,\bar{c}\sigma^{\lambda\tau}c
+\frac{1}{4}\mathcal{S}_{uu}^{\alpha}\gamma_{\alpha}d\,\bar{c}\gamma_{\mu}\gamma_{5}c\nonumber\\
&&-\frac{i}{4}\mathcal{S}_{uu}^{\alpha}\gamma_{\alpha}\sigma_{\lambda\mu}d\,\bar{c}\gamma^{\lambda}\gamma_{5}c
-\frac{i}{4}\mathcal{S}_{uu}^{\alpha}\gamma_{\alpha}\gamma_{\mu}d\,\bar{c}i\gamma_{5}c+2\left(S^\alpha_{uu} \to S^\alpha_{ud},\, d \to u \right)\, ,
\end{eqnarray}

\begin{eqnarray}
\sqrt{6}J^1_{\mu\nu}&=&-\frac{1}{4}\mathcal{S}_\mu^{uu}\gamma_{\nu}c\,\bar{c}d
+\frac{1}{4}\mathcal{S}_\mu^{uu}c\,\bar{c}\gamma_{\nu}d
-\frac{i}{4}\mathcal{S}_\mu^{uu}\sigma_{\lambda\nu}c\,\bar{c}\gamma^{\lambda}d
+\frac{1}{8}\mathcal{S}_\mu^{uu}\sigma_{\lambda\tau}\gamma_{\nu}c\,\bar{c}\sigma^{\lambda\tau}d\nonumber\\
&&-\frac{1}{4}\mathcal{S}_\mu^{uu}\gamma_{\lambda}\gamma_{\nu}\gamma_5c\,\bar{c}\gamma^{\lambda}\gamma_5d
-\frac{i}{4}\mathcal{S}_\mu^{uu}\gamma_{\nu}\gamma_{5}c\,\bar{c}i\gamma_{5}d
-\frac{1}{4}\mathcal{S}_\mu^{uu}\gamma_{\nu}d\,\bar{c}c
+\frac{1}{4}\mathcal{S}_\mu^{uu}d\,\bar{c}\gamma_{\nu}c\nonumber\\
&&-\frac{i}{4}\mathcal{S}_\mu^{uu}\sigma_{\lambda\nu}d\,\bar{c}\gamma^{\lambda}c
+\frac{1}{8}\mathcal{S}_\mu^{uu}\sigma_{\lambda\tau}\gamma_{\nu}d\,\bar{c}\sigma^{\lambda\tau}c
-\frac{1}{4}\mathcal{S}_\mu^{uu}\gamma_{\lambda}\gamma_{\nu}\gamma_5d\,\bar{c}\gamma^{\lambda}\gamma_5c\nonumber\\
&&-\frac{i}{4}\mathcal{S}_\mu^{uu}\gamma_{\nu}\gamma_{5}d\,\bar{c}i\gamma_{5}c+2\left(S_\mu^{uu} \to S_\mu^{ud},\, d \to u \right)
+\left(\mu \leftrightarrow \nu\right)\, ,
\end{eqnarray}

\begin{eqnarray}\label{Fierz-Jmunu2}
\sqrt{2}J^2_{\mu\nu}&=&-\frac{1}{4}\mathcal{S}_{ud}\gamma_5\gamma_{\nu}\gamma_{\mu}c\,\bar{c}u
+\frac{1}{4}\mathcal{S}_{ud}\gamma_5\gamma_{\nu}\gamma_{\lambda}\gamma_{\mu}c\,\bar{c}\gamma^{\lambda}u
+\frac{1}{8}\mathcal{S}_{ud}\gamma_5\gamma_{\nu}\sigma_{\lambda\tau}\gamma_{\mu}c\,\bar{c}\sigma^{\lambda\tau}u\nonumber\\
&&+\frac{1}{4}\mathcal{S}_{ud}\gamma_{\nu}\gamma_{\lambda}\gamma_{\mu}c\,\bar{c}\gamma^{\lambda}\gamma_5u
-\frac{i}{4}\mathcal{S}_{ud}\gamma_{\nu}\gamma_{\mu}c\,\bar{c}i\gamma_5 u -\frac{1}{4}\mathcal{S}_{ud}\gamma_5\gamma_{\nu}\gamma_{\mu}u\,\bar{c}c\nonumber\\
&&+\frac{1}{4}\mathcal{S}_{ud}\gamma_5\gamma_{\nu}\gamma_{\lambda}\gamma_{\mu}u\,\bar{c}\gamma^{\lambda}c
+\frac{1}{8}\mathcal{S}_{ud}\gamma_5\gamma_{\nu}\sigma_{\lambda\tau}\gamma_{\mu}u\,\bar{c}\sigma^{\lambda\tau}c
+\frac{1}{4}\mathcal{S}_{ud}\gamma_{\nu}\gamma_{\lambda}\gamma_{\mu}u\,\bar{c}\gamma^{\lambda}\gamma_5c\nonumber\\
&&-\frac{i}{4}\mathcal{S}_{ud}\gamma_{\nu}\gamma_{\mu}u\,\bar{c}i\gamma_5 c +\left(\mu\leftrightarrow \nu \right)\, ,
\end{eqnarray}
where $\mathcal{S}_{ud}\Gamma c=\varepsilon^{ijk}u^{Ti}C\gamma_5d^j\Gamma c^k$, $\mathcal{S}_{ud}\Gamma u=\varepsilon^{ijk}u^{Ti}C\gamma_5d^j\Gamma u^k$,
$\mathcal{S}^\mu_{uu}\Gamma c=\varepsilon^{ijk}u^{Ti}C\gamma^{\mu}u^j\Gamma c^k$,
$\mathcal{S}^\mu_{ud}\Gamma c=\varepsilon^{ijk}u^{Ti}C\gamma^{\mu}d^j\Gamma c^k$,
$\mathcal{S}^\mu_{uu}\Gamma d=\varepsilon^{ijk}u^{Ti}C\gamma^{\mu}u^j\Gamma d^k$,
$\mathcal{S}^\mu_{ud}\Gamma u=\varepsilon^{ijk}u^{Ti}C\gamma^{\mu}d^j\Gamma u^k$, the $\Gamma$ are Dirac matrixes.

The components $\mathcal{S}_{ud}\Gamma c$ and $\mathcal{S}_{ud}\Gamma u$ have the scalar diquark operator $\varepsilon^{ijk}u^{Ti}C\gamma_5d^j$, and can be classified as
the $\Lambda$-type currents, the components $\mathcal{S}^\mu_{uu}\Gamma c$,
$\mathcal{S}^\mu_{ud}\Gamma c$,
$\mathcal{S}^\mu_{uu}\Gamma d$ and
$\mathcal{S}^\mu_{ud}\Gamma u$  have the axialvector diquark operator $\varepsilon^{ijk}u^{Ti}C\gamma^{\mu} u^j$ or $\varepsilon^{ijk}u^{Ti}C\gamma^{\mu} d^j$, and  can be classified as
the $\Sigma$-type currents.
The components of the currents $J^1(x)$ and $J^2(x)$ have analogous $\Lambda$-type structures, while the components of the currents $J^3(x)$ and $J^4(x)$ have analogous $\Sigma$-type structures, the components of the currents $J_\mu^2(x)$, $J_\mu^3(x)$ and $J_\mu^4(x)$ have analogous $\Sigma$-type structures.
The currents have analogous  components  mix with each other potentially, however, the Fierz rearrangements (see Eqs.\eqref{Fierz-J1}-\eqref{Fierz-Jmunu2})  in the color and Dirac-spinor  spaces are not unique, which cannot exclude the mixings  between the $\Lambda$-type and $\Sigma$-type current operators if they have the same spin-parity $J^P$, direct calculations indicate that the non-diagonal correlation functions $\Pi^{ij}(p)$, $\Pi^{ij}_{\mu\nu}(p)$ and $\Pi^{ij}_{\mu\nu\alpha\beta}(p)\neq 0$ for $i\neq {j}$, where
\begin{eqnarray}
\Pi^{ij}(p)&=&i\int d^4x e^{ip \cdot x} \langle0|T\left\{J^{i}(x)\bar{J}^{j}(0)\right\}|0\rangle \, , \nonumber \\
\Pi^{ij}_{\mu\nu}(p)&=&i\int d^4x e^{ip \cdot x} \langle0|T\left\{J_{\mu}^{i}(x)\bar{J}_{\nu}^{j}(0)\right\}|0\rangle \, , \nonumber \\
\Pi^{ij}_{\mu\nu\alpha\beta}(p)&=&i\int d^4x e^{ip \cdot x} \langle0|T\left\{J_{\mu\nu}^{i}(x)\bar{J}^{j}_{\alpha\beta}(0)\right\}|0\rangle \, ,
\end{eqnarray}
the correlation functions shown in Eq.\eqref{CF-Pi-Pi-Pi} correspond to the case $i=j$.
We can introduce the mixing matrixes $U$, $J^{\prime i}=U_{ij}J^j$, $J_\mu^{\prime i}=U_{ij}J_\mu^j$ and $J_{\mu\nu}^{\prime i}=U_{ij}J_{\mu\nu}^j$, where the
$U$ are $4\times4$, $4\times4$ and $2\times2$ matrixes, respectively. Then we  obtain
the diagonal correlation functions,
\begin{eqnarray}
\Pi^{\prime ij}(p)&=& U_{im}\Pi^{ mn}(p)U^{\dagger}_{nj} \, , \nonumber \\
\Pi^{\prime ij}_{\mu\nu}(p)&=&U_{im}\Pi^{ mn}_{\mu\nu}(p)U^{\dagger}_{nj}\, , \nonumber \\
\Pi^{\prime ij}_{\mu\nu\alpha\beta}(p)&=&U_{im}\Pi^{ mn}_{\mu\nu\alpha\beta}(p)U^{\dagger}_{nj} \, ,
\end{eqnarray}
with the properties  $\Pi^{\prime ij}(p)$, $\Pi^{\prime ij}_{\mu\nu}(p)$, $\Pi^{\prime ij}_{\mu\nu\alpha\beta}(p)\propto \delta_{ij}$. The matrixes $U$ can be determined
by direct calculations based on the QCD sum rules, the tedious task may be our next work. The current operators $J^{\prime i}(x)$, $J_\mu^{\prime i}(x)$ and $J_{\mu\nu}^{\prime i}(x)$ couple potentially to more physical pentaquark states, which have more than one diquark-diquark-antiquark type Fock components.

 The color-singlet-color-singlet type current operators shown in Eqs.\eqref{Fierz-J1}-\eqref{Fierz-Jmunu2} couple potentially
   to the baryon-meson pairs or the pentaquark molecular states.
For example, the components $\mathcal{S}_{ud}c\,\bar{c}i\gamma_5 u $ and $\mathcal{S}_{ud}\gamma^\lambda\gamma_5u\,\bar{c}\gamma_{\lambda}c$ of the current $J^1(x)$ (also $J^2(x)$) couple potentially to the $\Lambda_c^+\bar{D}^0$ and $pJ/\psi$, respectively;
the components $\mathcal{S}^\mu_{uu}\gamma_{\mu}\gamma_{5}c\,\bar{c}i\gamma_{5}d$,
$\mathcal{S}^\mu_{ud}\gamma_{\mu}\gamma_{5}u\,\bar{c}i\gamma_{5}c$ and $\mathcal{S}^\mu_{ud}\gamma_{\lambda}\gamma_{\mu}u\,\bar{c}\gamma^{\lambda}c$ of the current $J^3(x)$ (also $J^4(x)$) couple potentially to the $\Sigma_c^{++}D^-$, $p \eta_c$ and $pJ/\psi$, respectively.
The diquark-diquark-antiquark type pentaquark  state can be taken as a special superposition of a series of  baryon-meson pairs or  pentaquark molecular states, and embodies  the net effects, the decays to its components (baryon-meson pairs) are Okubo-Zweig-Iizuka super-allowed.
From Eqs.\eqref{Fierz-J1}-\eqref{Fierz-Jmunu2}, we can see that there are   $\bar{c}c$, $\bar{c}i\gamma_5c$, $\bar{c}\gamma_{\mu}c$ and $\bar{c}\gamma_{\mu}\gamma_5c$
components in all the current operators, which have definite heavy quark spin,   the conversation of the heavy quark spin favors decay to the final states $\chi_{c0}$, $\eta_c$, $J/\psi$ and $\chi_{c1}$.

In fact, we should be careful  in performing  the Fierz rearrangements,  the rearrangements in the color and Dirac-spinor  spaces are non-trivial, the scenarios of the pentaquark states and molecular states are  different.
 The spatial separation among the diquark, diquark and antiquark leads  to small wave-function overlaps  to form the baryon-meson pairs, the rearrangements in the color and Dirac-spinor  spaces are  suppressed, which can account for the small widths of the $P_c(4312)$, $P_c(4440)$ and $P_c(4557)$ qualitatively.

It is difficult to take into account the non-local effects among the diquark, diquark and antiquark  in the currents   directly,
for example, the current $J^1(x)$ can be modified to
\begin{eqnarray}
J^1(x,\epsilon,\epsilon^\prime)&=&\varepsilon^{ila} \varepsilon^{ijk}\varepsilon^{lmn}  u^T_j(x+\epsilon) C\gamma_5 d_k(x+\epsilon)\,u^T_m(x) C\gamma_5 c_n(x)\,  C\bar{c}^{T}_{a}(x+\epsilon^\prime) \, ,
\end{eqnarray}
to account for the non-locality by adding two finite separations $\epsilon$ and $\epsilon^\prime$, but it is difficult to deal with the finite $\epsilon$ and $\epsilon^\prime$ in carrying out the operator product expansion, we have to take the limit $\epsilon, \, \epsilon^\prime \to 0$.

Now let us go back to the hadron representation of the correlation functions shown  in Eqs.\eqref{CF-Hadron-12}-\eqref{CF-Hadron-52}.  We obtain the spectral densities at the phenomenological side through  dispersion relation,
\begin{eqnarray}
\frac{{\rm Im}\Pi_{j}^1(s)}{\pi}&=& {\lambda^{-}_{j}}^2 \delta\left(s-M_{-}^2\right)+{\lambda^{+}_{j}}^2 \delta\left(s-M_{+}^2\right) =\, \rho^1_{j,H}(s) \, , \\
\frac{{\rm Im}\Pi^0_{j}(s)}{\pi}&=&M_{-}{\lambda^{-}_{j}}^2 \delta\left(s-M_{-}^2\right)-M_{+}{\lambda^{+}_{j}}^2 \delta\left(s-M_{+}^2\right)
=\rho^0_{j,H}(s) \, ,
\end{eqnarray}
where $j=\frac{1}{2}$, $\frac{3}{2}$, $\frac{5}{2}$, the subscript $H$ denotes  the hadron side,
then we introduce the  weight functions $\sqrt{s}\exp\left(-\frac{s}{T^2}\right)$ and $\exp\left(-\frac{s}{T^2}\right)$ to obtain the QCD sum rules
at the hadron side,
\begin{eqnarray}
\int_{4m_c^2}^{s_0}ds \left[\sqrt{s}\,\rho^1_{j,H}(s)+\rho^0_{j,H}(s)\right]\exp\left( -\frac{s}{T^2}\right)
&=&2M_{-}{\lambda^{-}_{j}}^2\exp\left( -\frac{M_{-}^2}{T^2}\right) \, ,
\end{eqnarray}
where the $s_0$ are the continuum threshold parameters, the $T^2$ are the Borel parameters.
We separate the  contributions  of the negative parity and positive parity pentaquark  states unambiguously.

\begin{figure}
 \centering
 \includegraphics[totalheight=3.0cm,width=15cm]{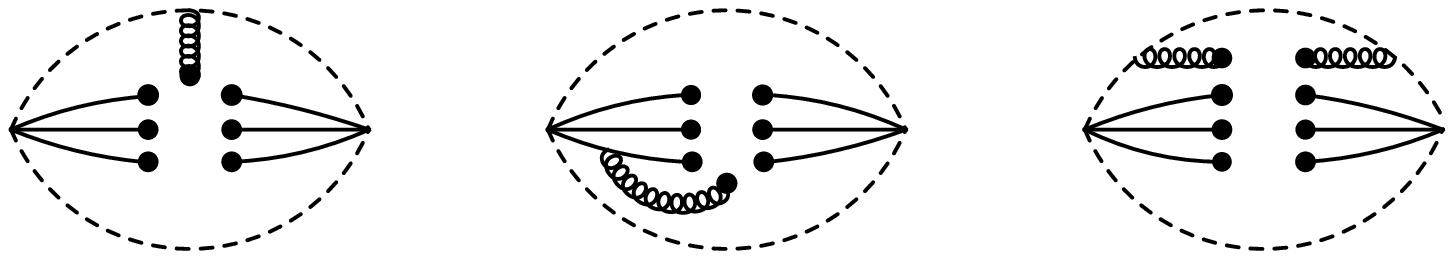}
  \vglue+3mm
 \includegraphics[totalheight=3.0cm,width=15cm]{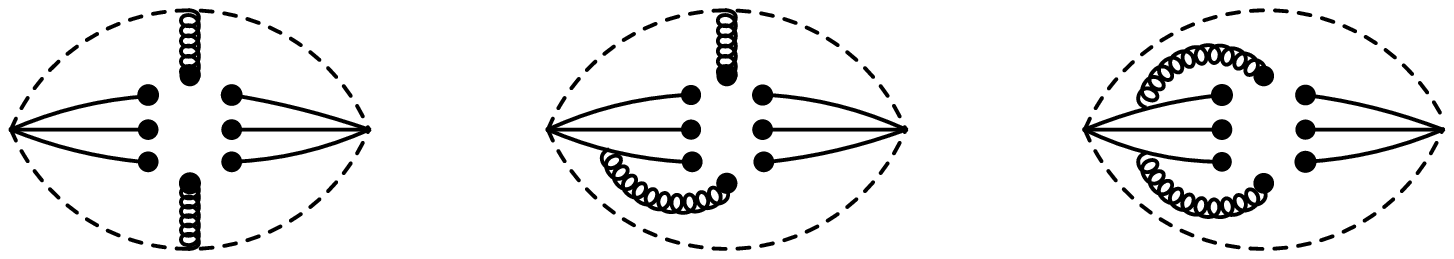}
    \caption{The diagrams contribute  to the  condensates $\langle\bar{q} q\rangle^2\langle\bar{q}g_s\sigma Gq\rangle $, $\langle\bar{q} q\rangle \langle\bar{q}g_s\sigma Gq\rangle^2 $, $\langle \bar{q}q\rangle^3\langle \frac{\alpha_s}{\pi}GG\rangle$. Other
diagrams obtained by interchanging of the $c$ quark lines (dashed lines) or light quark lines (solid lines) are implied. }\label{Feynman}
\end{figure}

In the following, we briefly outline  the operator product expansion for the correlation functions $\Pi(p)$, $\Pi_{\mu\nu}(p)$ and $\Pi_{\mu\nu\alpha\beta}(p)$ in perturbative QCD. Firstly,  we contract the $u$, $d$ and $c$ quark fields in the correlation functions $\Pi(p)$, $\Pi_{\mu\nu}(p)$ and $\Pi_{\mu\nu\alpha\beta}(p)$
 with Wick theorem, for example,
\begin{eqnarray}\label{QCD-Pi}
\Pi(p)&=&i\,\varepsilon^{ila}\varepsilon^{ijk}\varepsilon^{lmn}\varepsilon^{i^{\prime}l^{\prime}a^{\prime}}\varepsilon^{i^{\prime}j^{\prime}k^{\prime}}
\varepsilon^{l^{\prime}m^{\prime}n^{\prime}}\int d^4x e^{ip\cdot x} \nonumber\\
&&\Big\{  - Tr\Big[\gamma_5 D_{kk^\prime}(x) \gamma_5 C U^{T}_{jj^\prime}(x)C\Big] \,Tr\Big[\gamma_5 C_{nn^\prime}(x) \gamma_5 C U^{T}_{mm^\prime}(x)C\Big] C C_{a^{\prime}a}^T(-x)C   \nonumber\\
&&+  Tr \Big[\gamma_5 D_{kk^\prime}(x) \gamma_5 C U^{T}_{mj^\prime}(x)C \gamma_5 C_{nn^\prime}(x) \gamma_5 C U^{T}_{jm^\prime}(x)C\Big]   C C_{a^{\prime}a}^T(-x)C   \Big\} \, ,
\end{eqnarray}
for the current $J(x)=J^1(x)$, where
the $U_{ij}(x)$, $D_{ij}(x)$ and $C_{ij}(x)$ are the full $u$, $d$ and $c$ quark propagators respectively ($S_{ij}(x)=U_{ij}(x),\,D_{ij}(x)$),
 \begin{eqnarray}\label{L-quark-propagator}
S_{ij}(x)&=& \frac{i\delta_{ij}\!\not\!{x}}{ 2\pi^2x^4}-\frac{\delta_{ij}\langle
\bar{q}q\rangle}{12} -\frac{\delta_{ij}x^2\langle \bar{q}g_s\sigma Gq\rangle}{192} -\frac{ig_sG^{a}_{\alpha\beta}t^a_{ij}(\!\not\!{x}
\sigma^{\alpha\beta}+\sigma^{\alpha\beta} \!\not\!{x})}{32\pi^2x^2} \nonumber\\
&& -\frac{\delta_{ij}x^4\langle \bar{q}q \rangle\langle g_s^2 GG\rangle}{27648} -\frac{1}{8}\langle\bar{q}_j\sigma^{\mu\nu}q_i \rangle \sigma_{\mu\nu}+\cdots \, ,
\end{eqnarray}
\begin{eqnarray}\label{H-quark-propagator}
C_{ij}(x)&=&\frac{i}{(2\pi)^4}\int d^4k e^{-ik \cdot x} \left\{
\frac{\delta_{ij}}{\!\not\!{k}-m_c}
-\frac{g_sG^n_{\alpha\beta}t^n_{ij}}{4}\frac{\sigma^{\alpha\beta}(\!\not\!{k}+m_c)+(\!\not\!{k}+m_c)
\sigma^{\alpha\beta}}{(k^2-m_c^2)^2}\right.\nonumber\\
&&\left. -\frac{g_s^2 (t^at^b)_{ij} G^a_{\alpha\beta}G^b_{\mu\nu}(f^{\alpha\beta\mu\nu}+f^{\alpha\mu\beta\nu}+f^{\alpha\mu\nu\beta}) }{4(k^2-m_c^2)^5}+\cdots\right\} \, ,\nonumber\\
f^{\alpha\beta\mu\nu}&=&(\!\not\!{k}+m_c)\gamma^\alpha(\!\not\!{k}+m_c)\gamma^\beta(\!\not\!{k}+m_c)\gamma^\mu(\!\not\!{k}+m_c)\gamma^\nu(\!\not\!{k}+m_c)\, ,
\end{eqnarray}
and  $t^n=\frac{\lambda^n}{2}$, the $\lambda^n$ is the Gell-Mann matrix   \cite{PRT85}.  In Eq.\eqref{L-quark-propagator}, we retain the term $\langle\bar{q}_j\sigma_{\mu\nu}q_i \rangle$  comes from the Fierz re-arrangement of the $\langle q_i \bar{q}_j\rangle$ to  absorb the gluons  emitted from other  quark lines to form $\langle\bar{q}_j g_s G^a_{\alpha\beta} t^a_{mn}\sigma_{\mu\nu} q_i \rangle$    to extract the mixed condensate  $\langle\bar{q}g_s\sigma G q\rangle$ \cite{WangHuangTao}.  Then we compute  the integrals both in the coordinate space and momentum space to obtain the correlation functions $\Pi(p)$, $\Pi_{\mu\nu}(p)$ and $\Pi_{\mu\nu\alpha\beta}(p)$   at the quark level, and finally obtain the QCD spectral densities through   dispersion relation,
\begin{eqnarray}\label{QCD-rho1-2}
 \rho^1_{j,QCD}(s) &=&\frac{{\rm Im}\Pi_{j}^1(s)}{\pi}\, , \nonumber\\
\rho^0_{j,QCD}(s) &=&\frac{{\rm Im}\Pi_{j}^0(s)}{\pi}\, ,
\end{eqnarray}
where $j=\frac{1}{2}$, $\frac{3}{2}$, $\frac{5}{2}$. For more technical details, one can consult Ref.\cite{WangHuangTao}.
  In computing the integrals, we draw up all the Feynman diagrams from Eqs.\eqref{QCD-Pi}-\eqref{H-quark-propagator} and calculate them one by one.
In this article, we carry out the operator product expansion up to the vacuum condensates  of dimension-$13$ and assume vacuum saturation for the
 higher dimensional vacuum condensates. We take the truncations $n\leq 13$ and $k\leq 1$ in a consistent way,
the quark-gluon operators of the orders $\mathcal{O}( \alpha_s^{k})$ with $k> 1$ and dimension $n>13$ are  discarded.

In previous works \cite{Wang1508-EPJC,WangHuang-EPJC-1508-12,WangZG-EPJC-1509-12,WangZG-NPB-1512-32,WangZhang-APPB},  we took  the truncations $n\leq 10$ and $k\leq 1$
in the operator product expansion and discarded the quark-gluon operators of the orders $\mathcal{O}( \alpha_s^{k})$ with $k> 1$ and dimension $n>10$.  Sometimes  we also neglected the vacuum condensates   $\langle \frac{\alpha_sGG}{\pi}\rangle$,
 $\langle \bar{q}q\rangle\langle \frac{\alpha_sGG}{\pi}\rangle$, $\langle \bar{s}s\rangle\langle \frac{\alpha_sGG}{\pi}\rangle$, $\langle \bar{q}q\rangle^2\langle \frac{\alpha_sGG}{\pi}\rangle$, $\langle \bar{q}q\rangle \langle \bar{s}s\rangle\langle \frac{\alpha_sGG}{\pi}\rangle$,
 $ \langle \bar{s}s\rangle^2\langle \frac{\alpha_sGG}{\pi}\rangle$, which are not associated with the $\frac{1}{T^2}$, $\frac{1}{T^4}$ and $\frac{1}{T^6}$ to manifest themselves for the small Borel parameter $T^2$. We neglected those terms due to the small values of the gluon condensate $\langle \frac{\alpha_sGG}{\pi}\rangle$. In this article, we take into account all those contributions, such as $\langle \frac{\alpha_sGG}{\pi}\rangle$,
 $\langle \bar{q}q\rangle\langle \frac{\alpha_sGG}{\pi}\rangle$,  $\langle \bar{q}q\rangle^2\langle \frac{\alpha_sGG}{\pi}\rangle$.

 In this article, we re-examine the QCD side of the correlation functions $\Pi(p)$, $\Pi_{\mu\nu}(p)$ and $\Pi_{\mu\nu\alpha\beta}(p)$. From Eq.\eqref{QCD-Pi}, we can see that there are two $c$-quark propagators and three light quark propagators, if each $c$-quark line emits a gluon and each light quark line contributes  a quark-antiquark  pair, we obtain a operator $G_{\mu\nu}G_{\alpha\beta}\bar{u}u\bar{u}u\bar{d}d$, which is of dimension 13, see Fig.\ref{Feynman}.  We should take into account the vacuum condensates at least up to dimension $13$ in stead of dimension $10$.  The vacuum condensates  $\langle\bar{q} q\rangle^2\langle\bar{q}g_s\sigma Gq\rangle $, $\langle\bar{q} q\rangle \langle\bar{q}g_s\sigma Gq\rangle^2 $, $\langle \bar{q}q\rangle^3\langle \frac{\alpha_s}{\pi}GG\rangle$ are of dimension $11$ and $13$ respectively, and come from the Feynman diagrams shown in Fig.\ref{Feynman}.  Those vacuum condensates are  associated with the $\frac{1}{T^2}$, $\frac{1}{T^4}$ and $\frac{1}{T^6}$, which manifest themselves for  the small values of the $T^2$ and play an important role in determining the Borel windows, although at the Borel windows they play a minor important role.

 As we have obtained  the QCD spectral densities, see Eq.\eqref{QCD-rho1-2},  now  let us  match the hadron side with the QCD side of the correlation functions $\Pi(p)$, $\Pi_{\mu\nu}(p)$ and $\Pi_{\mu\nu\alpha\beta}(p)$, take the quark-hadron duality below the continuum thresholds  $s_0$, and  obtain  the  QCD sum rules:
\begin{eqnarray}\label{QCDSR}
2M_{-}\lambda^{-}_j{}^2\exp\left( -\frac{M_{-}^2}{T^2}\right)&=& \int_{4m_c^2}^{s_0}ds \,\rho_{QCD,j}(s)\,\exp\left( -\frac{s}{T^2}\right)\,  ,
\end{eqnarray}
where $\rho_{QCD,j}(s)=\sqrt{s}\rho_{QCD,j}^1(s)+\rho_{QCD,j}^{0}(s)$,
\begin{eqnarray}
\rho_{QCD,j}(s)&=&\rho^j_{0}(s)+\rho^j_{3}(s)+\rho^j_{4}(s)+\rho^j_{5}(s)+\rho^j_{6}(s)+\rho^j_{7}(s)+\rho^j_{8}(s)+\rho^j_{9}(s)+\rho^j_{10}(s)+\rho^j_{11}(s)\nonumber\\
&&+\rho^j_{13}(s)\, ,
\end{eqnarray}
\begin{eqnarray}
\rho^j_{0}(s)&\propto& {\rm perturbative \,\,\,\, terms}\, , \nonumber\\
\rho^j_{3}(s)&\propto& \langle \bar{q}q\rangle\, ,  \nonumber\\
\rho^j_{4}(s)&\propto& \langle \frac{\alpha_sGG}{\pi}\rangle\, ,  \nonumber\\
\rho^j_{5}(s)&\propto& \langle \bar{q}g_s\sigma Gq\rangle \, , \nonumber\\
\rho^j_{6}(s)&\propto& \langle \bar{q}q\rangle^2  \, , \nonumber\\
\rho^j_{7}(s)&\propto& \langle \bar{q}q\rangle\langle \frac{\alpha_sGG}{\pi}\rangle\, ,  \nonumber\\
\rho^j_{8}(s)&\propto& \langle\bar{q}q\rangle\langle \bar{q}g_s\sigma Gq\rangle\, ,   \nonumber\\
\rho^j_{9}(s)&\propto& \langle \bar{q}q\rangle^3\, ,  \nonumber\\
\rho^j_{10}(s)&\propto& \langle \bar{q}g_s\sigma Gq\rangle^2\, , \, \langle \bar{q}q\rangle^2\langle \frac{\alpha_sGG}{\pi}\rangle  \, , \nonumber\\
\rho^j_{11}(s)&\propto& \langle\bar{q}q\rangle^2\langle \bar{q}g_s\sigma Gq\rangle\, ,   \nonumber\\
\rho^j_{13}(s)&\propto& \langle\bar{q}q\rangle\langle \bar{q}g_s\sigma Gq\rangle^2\, , \, \langle \bar{q}q\rangle^3\langle \frac{\alpha_sGG}{\pi}\rangle  \, .
\end{eqnarray}
The explicit expressions of the QCD spectral densities are too lengthy to be presented here, the interested  reader  can obtain them by contacting  me via E-mail.

We derive   Eq.\eqref{QCDSR} with respect to  $\frac{1}{T^2}$, then eliminate the
 pole residues $\lambda^{-}_{j}$ and obtain the QCD sum rules for
 the masses of the hidden-charm  pentaquark states,
 \begin{eqnarray}
 M^2_{-} &=& \frac{-\int_{4m_c^2}^{s_0}ds \frac{d}{d(1/T^2)}\, \rho_{QCD,j}(s)\,\exp\left( -\frac{s}{T^2}\right)}{\int_{4m_c^2}^{s_0}ds \, \rho_{QCD,j}(s)\,\exp\left( -\frac{s}{T^2}\right)}\,  .
\end{eqnarray}

\section{Numerical results and discussions}
We take the vacuum condensates  to be the standard values
$\langle\bar{q}q \rangle=-(0.24\pm 0.01\, \rm{GeV})^3$,   $\langle\bar{q}g_s\sigma G q \rangle=m_0^2\langle \bar{q}q \rangle$,
$m_0^2=(0.8 \pm 0.1)\,\rm{GeV}^2$, $\langle \frac{\alpha_s
GG}{\pi}\rangle=(0.33\,\rm{GeV})^4 $    at the energy scale  $\mu=1\, \rm{GeV}$
\cite{SVZ79,PRT85,ColangeloReview}, and  take the $\overline{MS}$ mass $m_{c}(m_c)=(1.275\pm0.025)\,\rm{GeV}$
 from the Particle Data Group \cite{PDG}.
Moreover,  we take into account
the energy-scale dependence of  the quark condensate, mixed quark condensate and $\overline{MS}$ mass,
 \begin{eqnarray}
 \langle\bar{q}q \rangle(\mu)&=&\langle\bar{q}q \rangle({\rm 1GeV})\left[\frac{\alpha_{s}({\rm 1GeV})}{\alpha_{s}(\mu)}\right]^{\frac{12}{33-2n_f}}\, , \nonumber\\
 \langle\bar{q}g_s \sigma G q \rangle(\mu)&=&\langle\bar{q}g_s \sigma G q \rangle({\rm 1GeV})\left[\frac{\alpha_{s}({\rm 1GeV})}{\alpha_{s}(\mu)}\right]^{\frac{2}{33-2n_f}}\, ,\nonumber\\
m_c(\mu)&=&m_c(m_c)\left[\frac{\alpha_{s}(\mu)}{\alpha_{s}(m_c)}\right]^{\frac{12}{33-2n_f}} \, ,\nonumber\\
\alpha_s(\mu)&=&\frac{1}{b_0t}\left[1-\frac{b_1}{b_0^2}\frac{\log t}{t} +\frac{b_1^2(\log^2{t}-\log{t}-1)+b_0b_2}{b_0^4t^2}\right]\, ,
\end{eqnarray}
  where $t=\log \frac{\mu^2}{\Lambda^2}$, $b_0=\frac{33-2n_f}{12\pi}$, $b_1=\frac{153-19n_f}{24\pi^2}$, $b_2=\frac{2857-\frac{5033}{9}n_f+\frac{325}{27}n_f^2}{128\pi^3}$,  $\Lambda=210\,\rm{MeV}$, $292\,\rm{MeV}$  and  $332\,\rm{MeV}$ for the flavors  $n_f=5$, $4$ and $3$, respectively \cite{PDG,Narison-mix}, and evolve all the input parameters at the QCD side to the optimal  energy scales    $\mu$  with $n_f=4$ to extract the pentaquark masses.

In Refs.\cite{WangHbaryon-1,WangHbaryon-2,WangHbaryon-3,WangHbaryon-4,WangHbaryon-5,Wang-cc-baryon-penta},  we  study the  heavy, doubly-heavy and triply-heavy baryon states  with the QCD sum rules in a systematic way. In calculations, we observe that the continuum threshold parameters $\sqrt{s_0}=M_{gr}+ (0.5-0.8)\,\rm{GeV}$  work well,  where the subscript $gr$ denotes the ground state  baryon states.
The pentaquark states are another type baryon states due to  the fractional spins $1\over 2$, $3\over 2$, $5\over 2$. In the present work,  we take the continuum threshold parameters as $\sqrt{s_0}= M_{P}+(0.55-0.75)\,\rm{GeV}$.

 In this article, we choose the  Borel parameters $T^2$ and continuum threshold parameters $s_0$  to satisfy  the   four criteria:

$\bf 1.$ Pole dominance at the phenomenological side;

$\bf 2.$ Convergence of the operator product expansion;

$\bf 3.$ Appearance of the Borel platforms;

$\bf 4.$ Satisfying the energy scale formula,\\
via try and error.

Now we take a short digression to discuss the energy scale formula. The hidden-charm or hidden-bottom four-quark and five-quark systems   can be described
by a double-well potential in the heavy quark limit \cite{Wang1508-EPJC,WangHuang-EPJC-1508-12,WangZG-EPJC-1509-12,WangZG-NPB-1512-32,WangHuangTao,
Wang-tetra-formula-1,Wang-tetra-formula-2,Wang-tetra-IJMPA-1,Wang-tetra-IJMPA-2,Wang-tetra-IJMPA-3,WangHuang-molecule-1,WangHuang-molecule-2}. The heavy quark $Q$ serves as a  static well potential and  attracts   a light quark  to form a heavy diquark   in  color antitriplet $\bar{3}_c$. The heavy antiquark $\overline{Q}$ serves as another  static well potential and  attracts  a light antiquark or a light diquark  to form a heavy antidiquark or triquark  in  color triplet $3_c$. Then the diquark and antidiquark (or triquark) attract each other to form a compact tetraquark state (or pentaquark state).

The hidden-charm or hidden-bottom tetraquark states and pentaquark states are characterized by the effective $Q$-quark mass ${\mathbb{M}}_Q$ and the virtuality $V=\sqrt{M_{X/Y/Z/P}-(2{\mathbb{M}}_Q)^2}$ or the energy scale $\mu=\sqrt{M_{X/Y/Z/P}-(2{\mathbb{M}}_Q)^2}$ of the QCD spectral densities \cite{Wang1508-EPJC,WangHuang-EPJC-1508-12,WangZG-EPJC-1509-12,WangZG-NPB-1512-32,WangHuangTao,Wang-tetra-formula-1,Wang-tetra-formula-2,Wang-tetra-IJMPA-1,Wang-tetra-IJMPA-2,Wang-tetra-IJMPA-3}. The energy scale formula $\mu=\sqrt{M_{X/Y/Z/P}-(2{\mathbb{M}}_Q)^2}$ can enhance the pole contributions remarkably and improve the convergence of the operator product expansion considerably,   and works well in the QCD sum rules for the hidden-charm and hidden-bottom tetraquark states (hidden-charm pentaquark states).

In this article, we carry out the operator product expansion up to the vacuum condensates of dimension $13$, which is consistent with the dimension $10$ in the tetraquark case \cite{WangHuangTao,Wang-tetra-formula-1,Wang-tetra-formula-2,Wang-tetra-IJMPA-1,Wang-tetra-IJMPA-2,Wang-tetra-IJMPA-3},  and choose the updated value of the effective $c$-quark mass  ${\mathbb{M}}_c=1.82\,\rm{GeV}$ determined in the QCD sum rules for the hidden-charm  tetraquark states \cite{WangEPJC-1601}.
While in Refs.\cite{Wang1508-EPJC,WangHuang-EPJC-1508-12,WangZG-EPJC-1509-12,WangZG-NPB-1512-32}, we choose the old value ${\mathbb{M}}_c=1.80\,\rm{GeV}$.

In the following, let us go back to the Borel parameters and continuum threshold parameters.
After try and error, we obtain the  Borel parameters or Borel windows $T^2$, continuum threshold parameters $s_0$, ideal energy scales of the QCD spectral densities, pole contributions of the ground state pentaquark states, and contributions of the vacuum condensates of dimension $13$, which   are shown   explicitly in Table \ref{Borel}.

\begin{figure}
\centering
\includegraphics[totalheight=7cm,width=10cm]{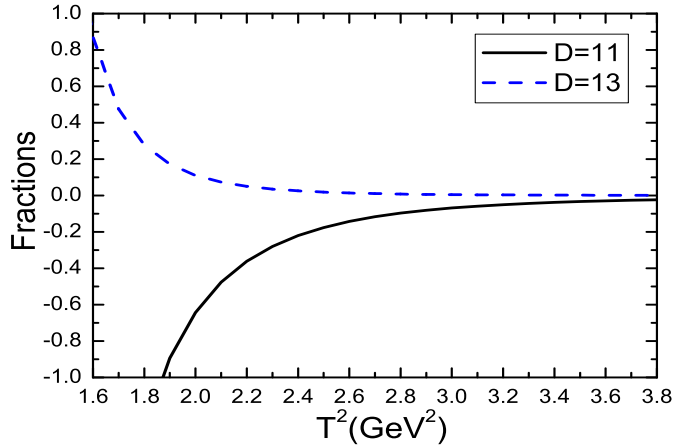}
  \caption{ The contributions of the vacuum condensates of dimension $11$ and $13$  with variation of the  Borel parameter $T^2$ for  the hidden-charm pentaquark state $[ud][uc]\bar{c}$ ($0$, $0$, $0$, $\frac{1}{2}$). }\label{fr-D13-fig}
\end{figure}

In Fig.\ref{fr-D13-fig}, we plot the contributions of the  vacuum condensates of dimension $11$ and $13$ (which were neglected in our previous works \cite{Wang1508-EPJC,WangHuang-EPJC-1508-12,WangZG-EPJC-1509-12,WangZG-NPB-1512-32}) with variation of the  Borel parameter $T^2$ for  the hidden-charm pentaquark state $[ud][uc]\bar{c}$ ($0$, $0$, $0$, $\frac{1}{2}$) with the central values of the parameters shown in Table \ref{Borel} as an example. From the figure, we can see that the vacuum condensates of dimension $13$ manifest themselves at the region $T^2< 2\,\rm{GeV}^2$, we should choose the value $T^2> 2\,\rm{GeV}^2$. On the other hand, the vacuum condensates of dimension $11$ manifest themselves at the region $T^2< 2.6\,\rm{GeV}^2$, which requires  a larger Borel parameter $T^2> 2.6\,\rm{GeV}^2$ to warrant the  convergence of the  operator product expansion.
The higher dimensional vacuum condensates play an important role in determining the Borel windows, we should take them into account in a consistent way, while in the Borel windows, they play an minor important role as the operator product expansion should be convergent, for example, in the present case, the contribution of the vacuum condensates of dimension $13$ is less than $1\%$, which is consistent with the analysis in Sect.2.

\begin{figure}
\centering
\includegraphics[totalheight=7cm,width=10cm]{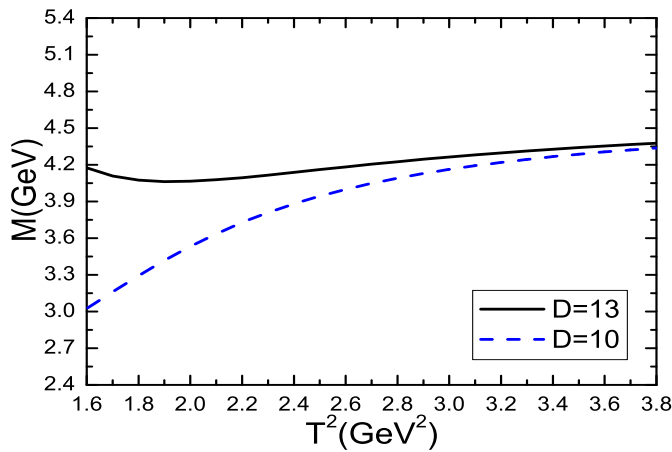}
  \caption{ The mass  with variation of the  Borel parameter $T^2$ for  the hidden-charm pentaquark state $[ud][uc]\bar{c}$ ($0$, $0$, $0$, $\frac{1}{2}$), the $D=10$, $13$ denote truncations of the operator product expansion. }\label{mass-D10-fig}
\end{figure}

In Fig.\ref{mass-D10-fig}, we plot the mass of the hidden-charm pentaquark state $[ud][uc]\bar{c}$ ($0$, $0$, $0$, $\frac{1}{2}$) with   variation of the  Borel parameter $T^2$ for truncations   of the operator product expansion up to the  vacuum condensates of dimension $10$ and $13$, respectively.   From the figure, we can see that
the vacuum condensates of dimension $11$ and $13$ play an important role to obtain stable QCD sum rules, we should take them into account.
In our previous works \cite{Wang1508-EPJC,WangHuang-EPJC-1508-12,WangZG-EPJC-1509-12,WangZG-NPB-1512-32}, we took into account the
 vacuum condensates up to dimension $10$ in carrying out the operator product expansion,  and sometimes neglected the vacuum condensates  $\langle\frac{\alpha_sGG}{\pi}\rangle$,
$\langle \bar{q}q\rangle\langle\frac{\alpha_sGG}{\pi}\rangle$ and $\langle \bar{q}q\rangle^2\langle\frac{\alpha_sGG}{\pi}\rangle$ due to their small contributions,
the Borel platforms were not flat enough. In the present work, we take into account the vacuum condensates up to dimension $13$ in a consistent way, and obtain very flat Borel platforms, the uncertainties originate from the Borel parameters are tiny.

From the Table \ref{Borel}, we can see that the pole contributions are about $(40-60)\%$ and the contributions of the vacuum condensates of dimension $13$ are $\leq 1\%$ or $\ll 1\%$,  the pole dominance  at the hadron side is satisfied and  the operator product expansion is well convergent,  the first two criteria or the two basic criteria of the QCD sum rules are satisfied, so we expect to make reasonable predictions.

We take into account  all uncertainties  of the input   parameters,
and obtain  the masses and pole residues of
 the negative parity   hidden-charm pentaquark states, which are shown explicitly in Table \ref{mass}. From Table \ref{Borel} and Table \ref{mass}, we can see that the energy scale formula $\mu=\sqrt{M_{P}-(2{\mathbb{M}}_c)^2}$ is satisfied, the  criterion  $\bf 4$ is  satisfied.

\begin{figure}
\centering
\includegraphics[totalheight=6cm,width=7cm]{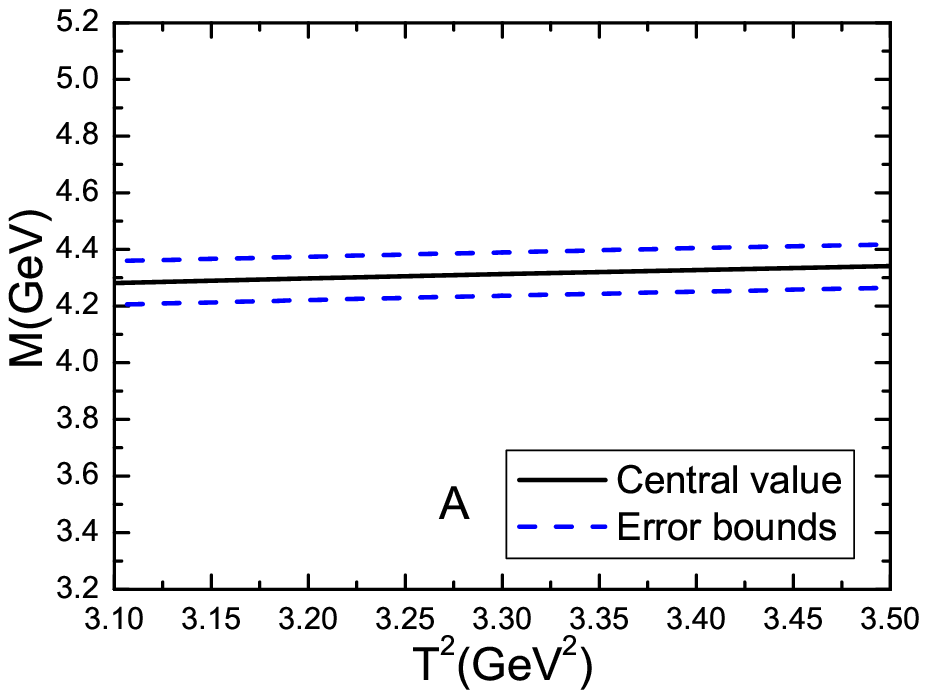}
\includegraphics[totalheight=6cm,width=7cm]{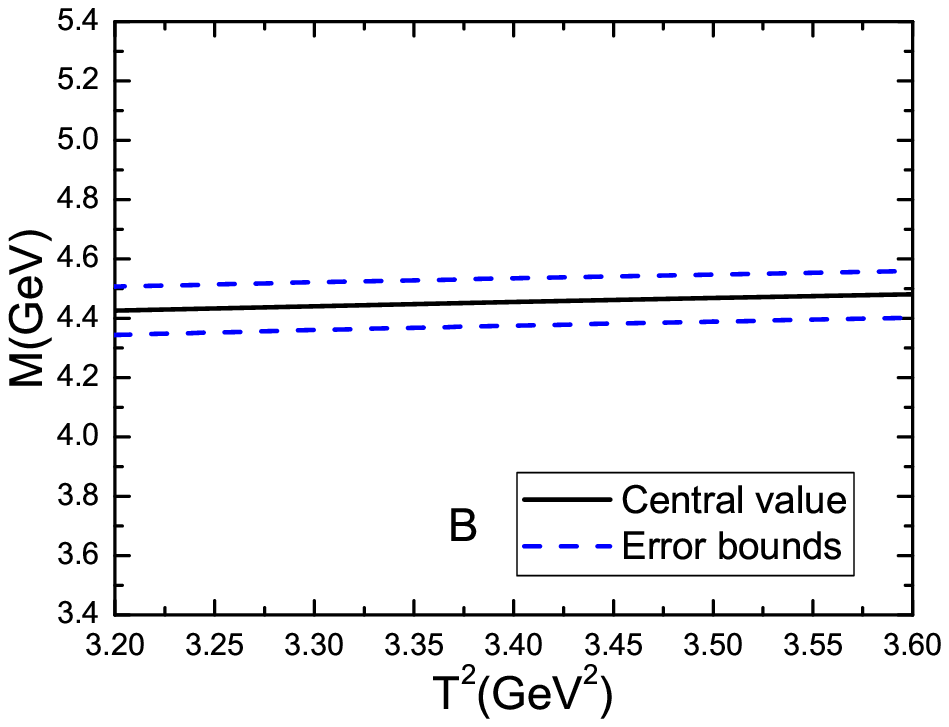}
\includegraphics[totalheight=6cm,width=7cm]{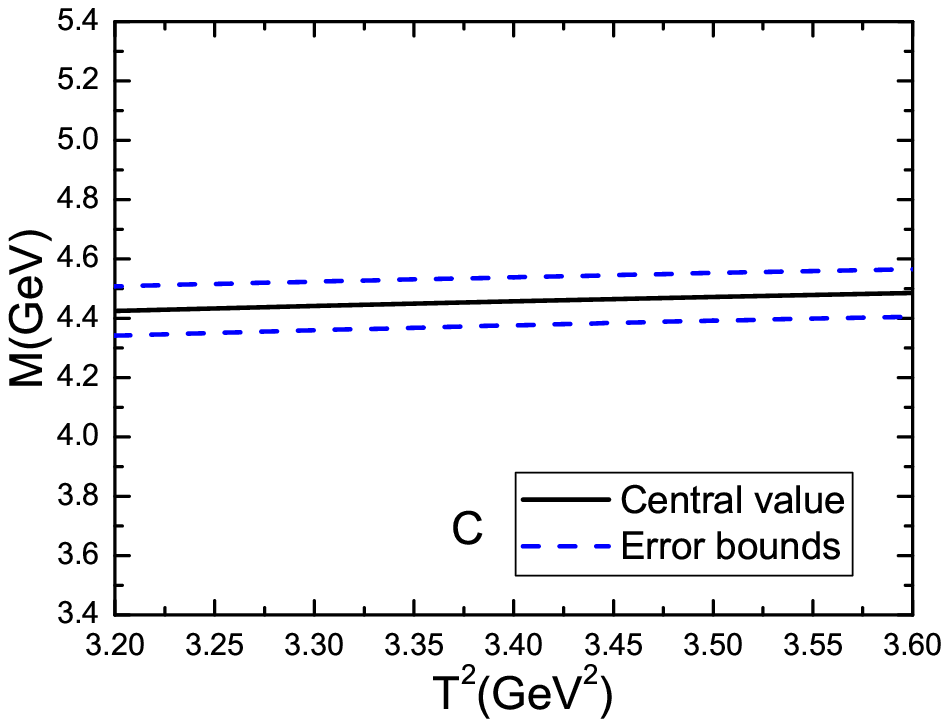}
\includegraphics[totalheight=6cm,width=7cm]{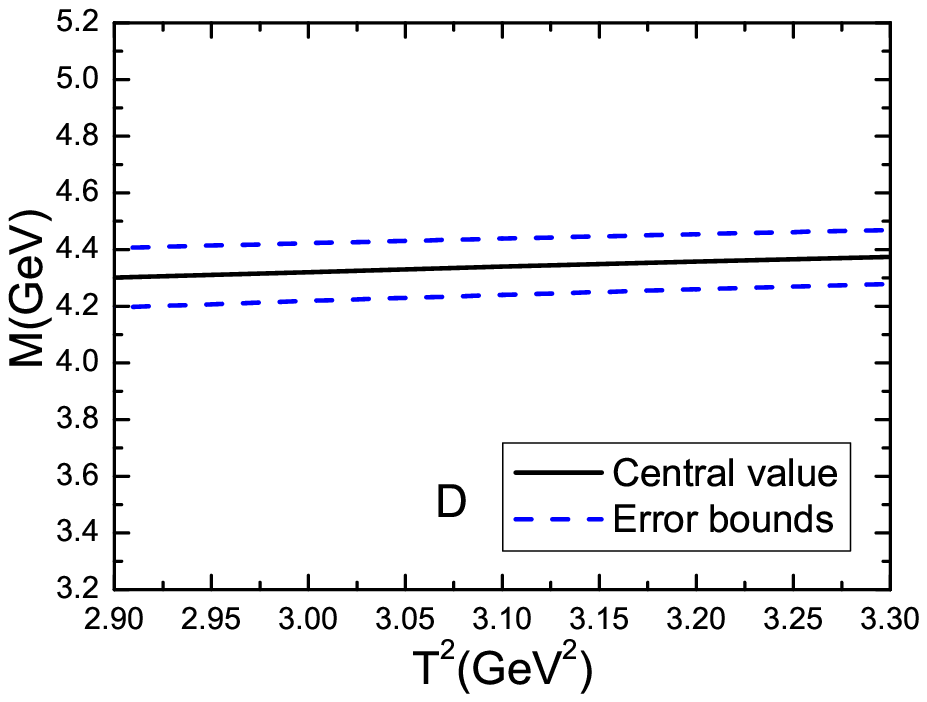}
\includegraphics[totalheight=6cm,width=7cm]{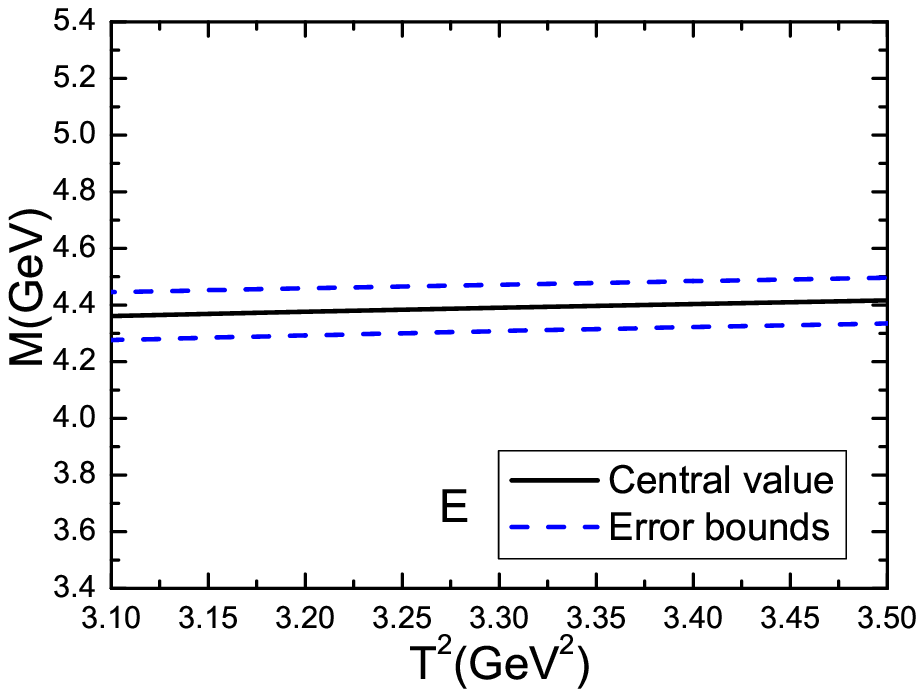}
\includegraphics[totalheight=6cm,width=7cm]{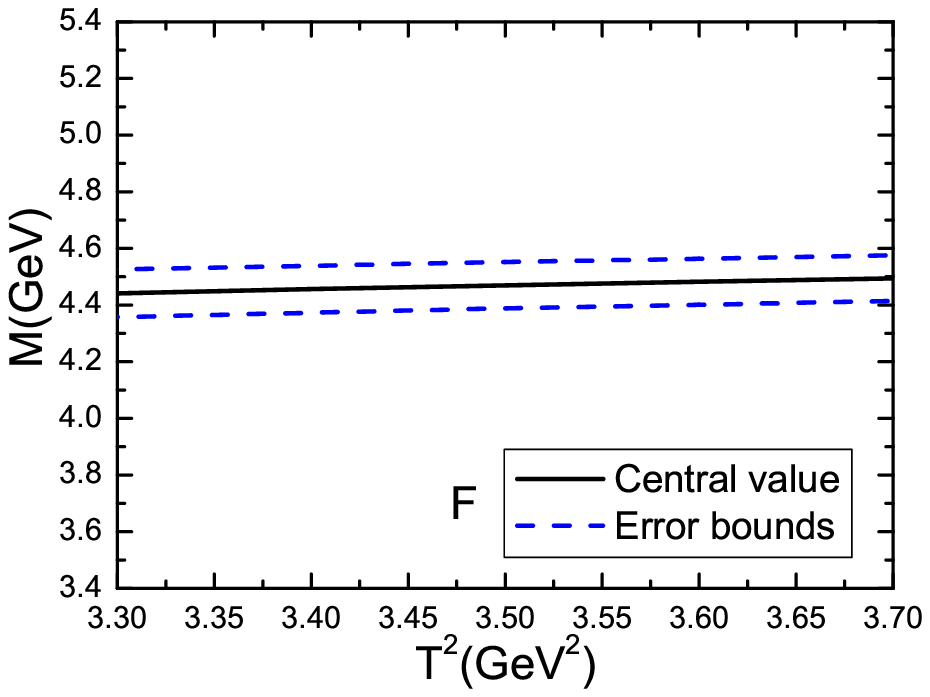}
  \caption{ The masses  with variations of the  Borel parameters $T^2$ for  the hidden-charm pentaquark states, the $A$, $B$, $C$, $D$, $E$ and $F$  denote the
  pentaquark states   $[ud][uc]\bar{c}$ ($0$, $0$, $0$, $\frac{1}{2}$),
$[ud][uc]\bar{c}$ ($0$, $1$, $1$, $\frac{1}{2}$),
$[uu][dc]\bar{c}+2[ud][uc]\bar{c}$ ($1$, $0$, $1$, $\frac{1}{2}$),
$[uu][dc]\bar{c}+2[ud][uc]\bar{c}$ ($1$, $1$, $0$, $\frac{1}{2}$),
$[ud][uc]\bar{c}$ ($0$, $1$, $1$, $\frac{3}{2}$) and $[uu][dc]\bar{c}+2[ud][uc]\bar{c}$ ($1$, $0$, $1$, $\frac{3}{2}$), respectively. }\label{mass-1-fig}
\end{figure}

\begin{figure}
\centering
\includegraphics[totalheight=6cm,width=7cm]{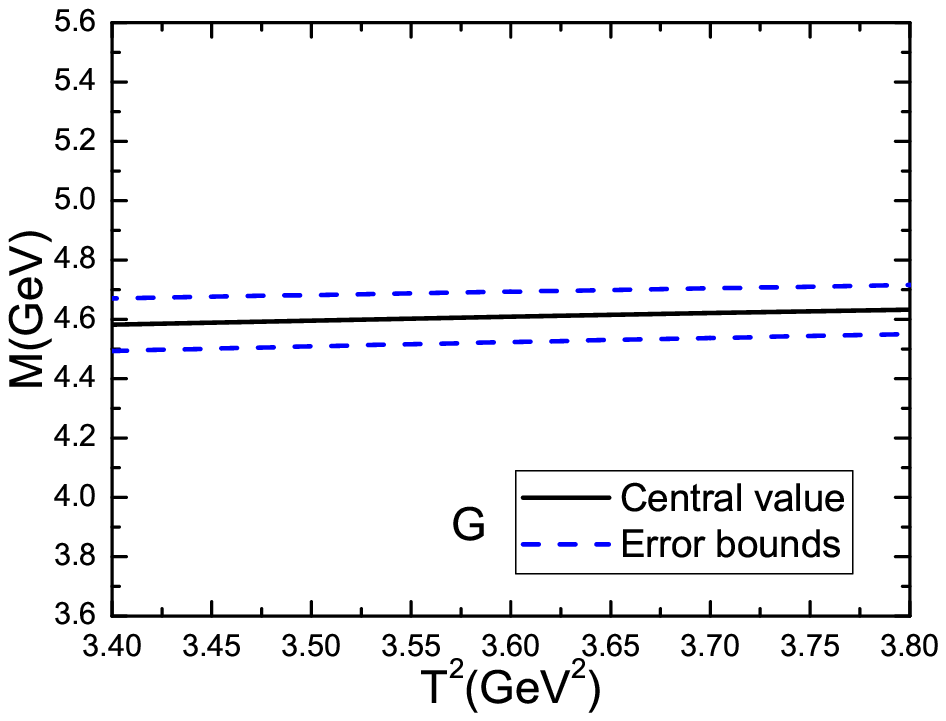}
\includegraphics[totalheight=6cm,width=7cm]{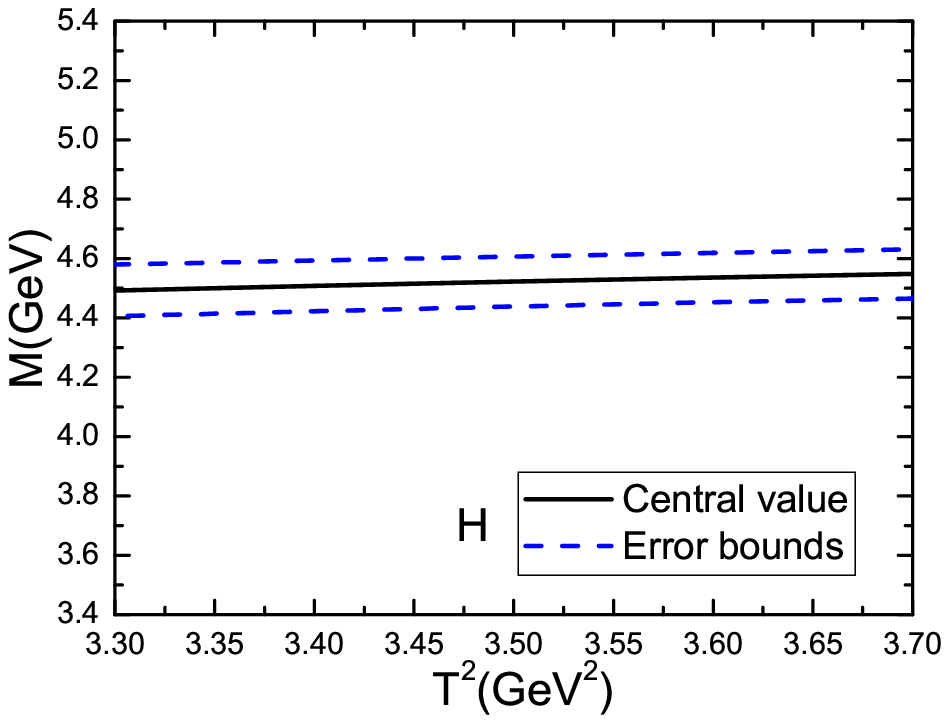}
\includegraphics[totalheight=6cm,width=7cm]{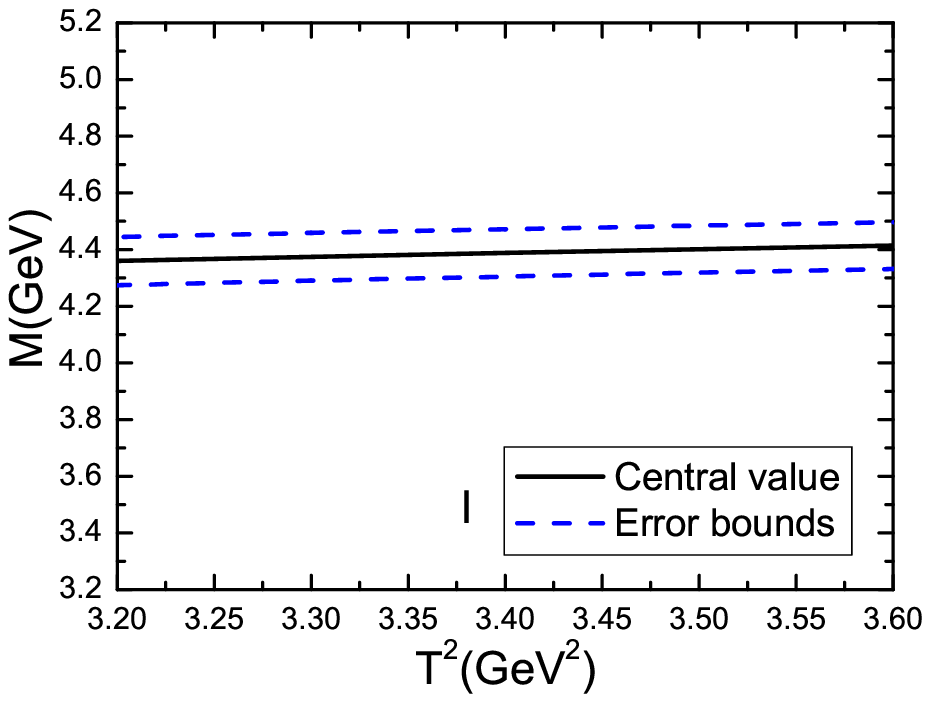}
\includegraphics[totalheight=6cm,width=7cm]{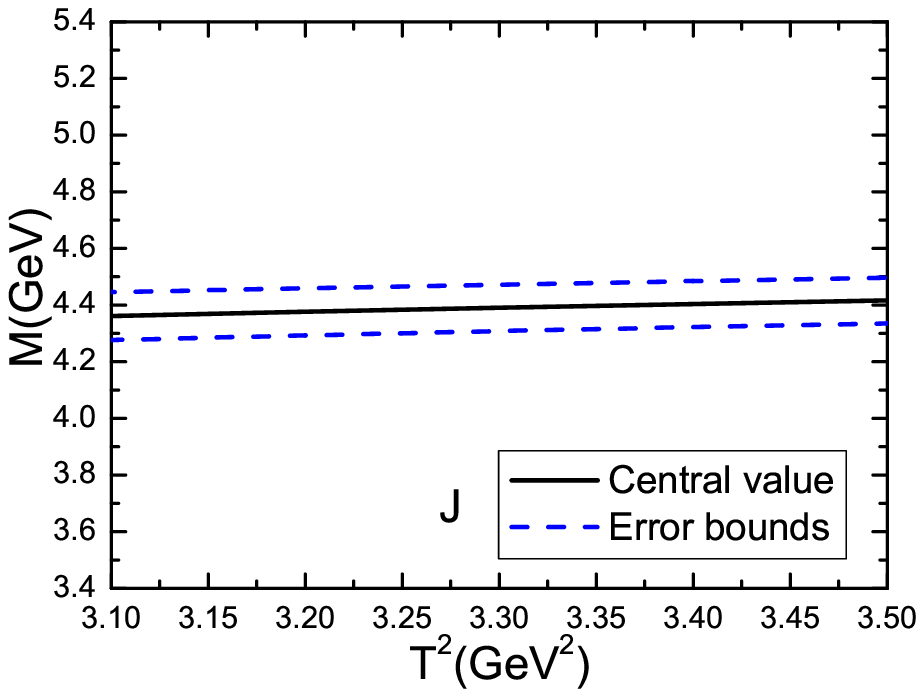}
  \caption{ The masses  with variations of the  Borel parameters $T^2$ for  the hidden-charm pentaquark states, the $G$, $H$, $I$  and $J$  denote the
   pentaquark states  $[uu][dc]\bar{c}+2[ud][uc]\bar{c}$ ($1$, $1$, $2$, $\frac{3}{2}$),
$[uu][dc]\bar{c}+2[ud][uc]\bar{c}$ ($1$, $1$, $2$, $\frac{3}{2}$),
$[uu][dc]\bar{c}+2[ud][uc]\bar{c}$ ($1$, $1$, $2$, $\frac{5}{2}$) and
$[ud][uc]\bar{c}$ ($0$, $1$, $1$, $\frac{5}{2}$), respectively. }\label{mass-2-fig}
\end{figure}

 In Figs.\ref{mass-1-fig}-\ref{mass-2-fig}, we plot the   masses  of the hidden-charm pentaquark states   with variations of the Borel parameters $T^2$ in  the Borel windows. From the figures, we can see that there appear very flat platforms, the    criterion  $\bf 3$ is  satisfied. Now the four criteria of the QCD sum rules are all satisfied, we expect to make robust predictions.

From Table \ref{mass}, we can see that the mass-splittings among those $J^{P}={\frac{1}{2}}^-$, ${\frac{3}{2}}^-$ and ${\frac{5}{2}}^-$ pentaquark states are rather small, about or less than $0.3\,\rm{GeV}$. In this article, we take the scalar and axialvector diquark  states as the basic constituents to study the pentaquark states. The calculations based on the QCD sum rules indicate that the light axialvector diquark states $\varepsilon^{ijk}u^{Ti}C\gamma_{\mu} u^j$ and $\varepsilon^{ijk}u^{Ti}C\gamma_{\mu} d^j$  have  a larger mass than the corresponding the scalar diquark state $\varepsilon^{ijk}u^{Ti}C\gamma_{5} d^j$, about $0.15-0.20\,\rm{GeV}$ \cite{WangLightDiquark}, while the heavy scalar and axialvector diquark states $\varepsilon^{ijk}q^{Ti}C\gamma_{\mu} c^j$ and $\varepsilon^{ijk}q^{Ti}C\gamma_{5} c^j$ have almost degenerated masses \cite{WangDiquark-1,WangDiquark-2}. In this way, we can account for the  small pentaquark mass splittings reasonably. In fact, the QCD calculations differ from
 quark model calculations significantly, the pentaquark masses shown in Table \ref{mass}  are not directly related to the diquark masses,  we  obtain them with the full QCD sum rules by imposing the same criteria.

The predicted masses $M_{P}=4.31\pm0.11\,\rm{GeV}$ for the ground state $[ud][uc]\bar{c}$ ($0$, $0$, $0$, $\frac{1}{2}$) pentaquark state and
$M_{P}=4.34\pm0.14\,\rm{GeV}$ for the ground state $[uu][dc]\bar{c}+2[ud][uc]\bar{c}$ ($1$, $1$, $0$, $\frac{1}{2}$) pentaquark state
are both in excellent agreement  with the experimental data   $M_{P(4312)}=4311.9\pm0.7^{+6.8}_{-0.6} \,\rm{MeV}$ from the LHCb   collaboration \cite{LHCb-Pc4312}, and support assigning the $P_c(4312)$ to be the hidden-charm pentaquark state with $J^{P}={\frac{1}{2}}^-$.

The predicted masses
$M_{P}=4.45\pm0.11\,\rm{GeV}$ for the ground state $[ud][uc]\bar{c}$ ($0$, $1$, $1$, $\frac{1}{2}$)  pentaquark state,
$M_{P}=4.46\pm0.11\,\rm{GeV}$ for the ground state $[uu][dc]\bar{c}+2[ud][uc]\bar{c}$ ($1$, $0$, $1$, $\frac{1}{2}$)  pentaquark state and
$M_{P}=4.39\pm0.11$ for the ground state $[ud][uc]\bar{c}$ ($0$, $1$, $1$, $\frac{3}{2}$), $[uu][dc]\bar{c}+2[ud][uc]\bar{c}$ ($1$, $1$, $2$, $\frac{5}{2}$), $[ud][uc]\bar{c}$ ($0$, $1$, $1$, $\frac{5}{2}$)    pentaquark states
are in excellent agreement (or  compatible with) the experimental data   $M_{P(4440)}=4440.3\pm1.3^{+4.1}_{-4.7} \,\rm{MeV}$ from the LHCb   collaboration \cite{LHCb-Pc4312}, and support assigning the $P_c(4440)$ to be the hidden-charm pentaquark state with $J^{P}={\frac{1}{2}}^-$, ${\frac{3}{2}}^-$ or ${\frac{5}{2}}^-$.

  The predicted masses
$M_{P}=4.45\pm0.11\,\rm{GeV}$ for the ground state $[ud][uc]\bar{c}$ ($0$, $1$, $1$, $\frac{1}{2}$)  pentaquark state,
$M_{P}=4.46\pm0.11\,\rm{GeV}$ for the ground state $[uu][dc]\bar{c}+2[ud][uc]\bar{c}$ ($1$, $0$, $1$, $\frac{1}{2}$)  pentaquark state   and $M_{P}=4.47\pm0.11\,\rm{GeV}$ for the ground state
 $[uu][dc]\bar{c}+2[ud][uc]\bar{c}$ ($1$, $0$, $1$, $\frac{3}{2}$)    pentaquark states
are in excellent agreement  the experimental data   $M_{P(4457)}=4457.3\pm0.6^{+4.1}_{-1.7} \,\rm{MeV}$ from the LHCb   collaboration \cite{LHCb-Pc4312}, and support assigning the $P_c(4457)$ to be the hidden-charm pentaquark state with $J^{P}={\frac{1}{2}}^-$ or ${\frac{3}{2}}^-$.

In Table \ref{mass}, we present the possible assignments of the $P_c(4312)$, $P_c(4440)$ and $P_c(4457)$ explicitly as a summary. In Table \ref{mass-1508-et al}, we compare the present predictions with our previous calculations \cite{Wang1508-EPJC,WangHuang-EPJC-1508-12,WangZG-EPJC-1509-12,WangZG-NPB-1512-32}, where the vacuum condensates of dimension $11$ and $13$ were  neglected, sometimes the vacuum condensates $\langle\frac{\alpha_sGG}{\pi}\rangle$,
$\langle \bar{q}q\rangle\langle\frac{\alpha_sGG}{\pi}\rangle$ and $\langle \bar{q}q\rangle^2\langle\frac{\alpha_sGG}{\pi}\rangle$ were also neglected. From the Table \ref{mass-1508-et al}, we can see that in some cases the predicted masses change  remarkably, while in other cases the predicted masses change  slightly. All in all, the uncertainties of the present  pentaquark masses are smaller than the corresponding  old ones, as we obtain more flat Borel platforms in the present work.

\begin{table}
\begin{center}
\begin{tabular}{|c|c|c|c|c|c|c|c|}\hline\hline
                  &$T^2 \rm{GeV}^2)$     &$\sqrt{s_0}(\rm{GeV})$    &$\mu(\rm{GeV})$  &pole          &$D_{13}$         \\ \hline

$J^1(x)$          &$3.1-3.5$             &$4.96\pm0.10$             &$2.3$            &$(41-62)\%$   &$<1\%$      \\ \hline

$J^2(x)$          &$3.2-3.6$             &$5.10\pm0.10$             &$2.6$            &$(42-63)\%$   &$<1\%$      \\ \hline

$J^3(x)$          &$3.2-3.6$             &$5.11\pm0.10$             &$2.6$            &$(42-63)\%$   &$\ll1\%$      \\ \hline

$J^4(x)$          &$2.9-3.3$             &$5.00\pm0.10$             &$2.4$            &$(40-64)\%$   &$\leq1\%$     \\ \hline

$J^1_\mu(x)$      &$3.1-3.5$             &$5.03\pm0.10$             &$2.4$            &$(42-63)\%$   &$\leq1\%$     \\ \hline

$J^2_\mu(x)$      &$3.3-3.7$             &$5.11\pm0.10$             &$2.6$            &$(40-61)\%$   &$\ll1\%$     \\ \hline

$J^3_\mu(x)$      &$3.4-3.8$             &$5.26\pm0.10$             &$2.8$            &$(42-62)\%$   &$\ll1\%$     \\ \hline

$J^4_\mu(x)$      &$3.3-3.7$             &$5.17\pm0.10$             &$2.7$            &$(41-61)\%$   &$<1\%$     \\ \hline

$J^1_{\mu\nu}(x)$ &$3.2-3.6$             &$5.03\pm0.10$             &$2.4$            &$(40-61)\%$   &$\leq1\%$     \\ \hline

$J^2_{\mu\nu}(x)$ &$3.1-3.5$             &$5.03\pm0.10$             &$2.4$            &$(42-63)\%$   &$\leq1\%$     \\ \hline\hline
\end{tabular}
\end{center}
\caption{ The Borel  windows, continuum threshold parameters, ideal energy scales, pole contributions,   contributions of the vacuum condensates of dimension 13 for the hidden-charm pentaquark states. }\label{Borel}
\end{table}

\begin{table}
\begin{center}
\begin{tabular}{|c|c|c|c|c|c|c|c|c|}\hline\hline
$[qq^\prime][q^{\prime\prime}c]\bar{c}$ ($S_L$, $S_H$, $J_{LH}$, $J$) &$M(\rm{GeV})$   &$\lambda(10^{-3}\rm{GeV}^6)$ &Assignments        &Currents \\ \hline

$[ud][uc]\bar{c}$ ($0$, $0$, $0$, $\frac{1}{2}$)                      &$4.31\pm0.11$   &$1.40\pm0.23$                &$?\,P_c(4312)$      &$J^1(x)$    \\

$[ud][uc]\bar{c}$ ($0$, $1$, $1$, $\frac{1}{2}$)                      &$4.45\pm0.11$   &$3.02\pm0.48$                &$?\,P_c(4440/4457)$ &$J^2(x)$    \\

$[uu][dc]\bar{c}+2[ud][uc]\bar{c}$ ($1$, $0$, $1$, $\frac{1}{2}$)     &$4.46\pm0.11$   &$4.32\pm0.71$                &$?\,P_c(4440/4457)$ &$J^3(x)$    \\

$[uu][dc]\bar{c}+2[ud][uc]\bar{c}$ ($1$, $1$, $0$, $\frac{1}{2}$)     &$4.34\pm0.14$   &$3.23\pm0.61$                &$?\,P_c(4312)$      &$J^4(x)$   \\

$[ud][uc]\bar{c}$ ($0$, $1$, $1$, $\frac{3}{2}$)                      &$4.39\pm0.11$   &$1.44\pm0.23$                &$?\,P_c(4440)$      &$J^1_\mu(x)$ \\

$[uu][dc]\bar{c}+2[ud][uc]\bar{c}$ ($1$, $0$, $1$, $\frac{3}{2}$)     &$4.47\pm0.11$   &$2.41\pm0.38$                &$?\,P_c(4440/4457)$ &$J^2_\mu(x)$  \\

$[uu][dc]\bar{c}+2[ud][uc]\bar{c}$ ($1$, $1$, $2$, $\frac{3}{2}$)     &$4.61\pm0.11$   &$5.13\pm0.79$                &                    &$J^3_\mu(x)$  \\

$[uu][dc]\bar{c}+2[ud][uc]\bar{c}$ ($1$, $1$, $2$, $\frac{3}{2}$)     &$4.52\pm0.11$   &$4.49\pm0.72$                &                    &$J^4_\mu(x)$ \\

$[uu][dc]\bar{c}+2[ud][uc]\bar{c}$ ($1$, $1$, $2$, $\frac{5}{2}$)     &$4.39\pm0.11$   &$1.94\pm0.31$                &$?\,P_c(4440)$      &$J^1_{\mu\nu}(x)$ \\

$[ud][uc]\bar{c}$ ($0$, $1$, $1$, $\frac{5}{2}$)                      &$4.39\pm0.11$   &$1.44\pm0.23$                &$?\,P_c(4440)$      &$J^2_{\mu\nu}(x)$ \\ \hline\hline
\end{tabular}
\end{center}
\caption{ The masses  and pole residues of the hidden-charm pentaquark states. }\label{mass}
\end{table}

\begin{table}
\begin{center}
\begin{tabular}{|c|c|c|c|c|c|c|c|c|}\hline\hline
$[qq^\prime][q^{\prime\prime}c]\bar{c}$ ($S_L$, $S_H$, $J_{LH}$, $J$) &This work       &Previous work                 &Currents \\ \hline

$[ud][uc]\bar{c}$ ($0$, $0$, $0$, $\frac{1}{2}$)                      &$4.31\pm0.11$   &$4.29\pm 0.13$                &$J^1(x)$    \\

$[ud][uc]\bar{c}$ ($0$, $1$, $1$, $\frac{1}{2}$)                      &$4.45\pm0.11$   &$4.30 \pm 0.13$               &$J^2(x)$    \\

$[uu][dc]\bar{c}+2[ud][uc]\bar{c}$ ($1$, $0$, $1$, $\frac{1}{2}$)     &$4.46\pm0.11$   &$4.42 \pm 0.12$               &$J^3(x)$    \\

$[uu][dc]\bar{c}+2[ud][uc]\bar{c}$ ($1$, $1$, $0$, $\frac{1}{2}$)     &$4.34\pm0.14$   &$4.35\pm 0.15$                &$J^4(x)$   \\

$[ud][uc]\bar{c}$ ($0$, $1$, $1$, $\frac{3}{2}$)                      &$4.39\pm0.11$   &$4.38 \pm 0.13$               &$J^1_\mu(x)$ \\

$[uu][dc]\bar{c}+2[ud][uc]\bar{c}$ ($1$, $0$, $1$, $\frac{3}{2}$)     &$4.47\pm0.11$   &$4.39\pm 0.13$                &$J^2_\mu(x)$  \\

$[uu][dc]\bar{c}+2[ud][uc]\bar{c}$ ($1$, $1$, $2$, $\frac{3}{2}$)     &$4.61\pm0.11$   &$4.39 \pm 0.14$               &$J^3_\mu(x)$  \\

$[uu][dc]\bar{c}+2[ud][uc]\bar{c}$ ($1$, $1$, $2$, $\frac{3}{2}$)     &$4.52\pm0.11$   &$4.39 \pm 0.14$               &$J^4_\mu(x)$ \\

$[uu][dc]\bar{c}+2[ud][uc]\bar{c}$ ($1$, $1$, $2$, $\frac{5}{2}$)     &$4.39\pm0.11$   &                              &$J^1_{\mu\nu}(x)$ \\

$[ud][uc]\bar{c}$ ($0$, $1$, $1$, $\frac{5}{2}$)                      &$4.39\pm0.11$   &                              &$J^2_{\mu\nu}(x)$ \\ \hline\hline
\end{tabular}
\end{center}
\caption{ The masses (in unit of GeV) are compared with the old calculations in our previous works \cite{Wang1508-EPJC,WangHuang-EPJC-1508-12,WangZG-EPJC-1509-12,WangZG-NPB-1512-32}.   }\label{mass-1508-et al}
\end{table}

The
diquark-diquark-antiquark type pentaquark  state can be taken as a special superposition of a series of  baryon-meson pairs or  pentaquark molecular states, and embodies  the net effects, the decays to its components (baryon-meson pairs) are Okubo-Zweig-Iizuka super-allowed. We can study the two-body strong decays of the pentaquark states exclusively with the three-point QCD sum rules with the guideline of the Fierz rearrangements in Eqs.\eqref{Fierz-J1}-\eqref{Fierz-Jmunu2},
\begin{eqnarray}
P_{c} &\to& p J/\psi  \, , \,p\eta_c \, , \,p\chi_{c0} \, , \,p\chi_{c1}\, , \,\Delta J/\psi  \, , \,\Delta\eta_c \, , \, N(1440)J/\psi \, , \,N(1440)\eta_c \, , \, \Lambda_c \bar{D}^*\, , \, \Lambda_c \bar{D}\, , \,  \nonumber\\
&& \Lambda_c(2595) \bar{D}^{*}\, , \,  \Lambda_c(2595) \bar{D}\, , \, \Sigma_c \bar{D}\, , \, \Sigma_c \bar{D}^*\, , \, \Sigma_c^* \bar{D}\, , \,\Sigma_c^* \bar{D}^*\, .
\end{eqnarray}
It is better to use the partial decay widths and total width  besides the mass to assign or distinguish   a pentaquark candidate. We can search for those hidden-charm pentaquark states in the $J/\psi p$, $p\eta_c$, $\cdots$,
$\Sigma_c^* \bar{D}^{*}$  invariant mass distributions and confront  the present predictions  to the experimental data in the future as the first step.

In Ref.\cite{WangPenta-IJMPA}, we choose the component $\mathcal{S}^\alpha_{ud}\gamma_{\alpha}\gamma_{5}c\,\bar{c}i\gamma_{5}u$ in the currents $J^3(x)$ and $J^4(x)$ to interpolate the $\Sigma\bar{D}$ pentaquark molecular state;
choose the component $\mathcal{S}_\mu^{ud}c\,\bar{c}i\gamma_{5}u$ in the currents $J^2_\mu(x)$ and $J^3_\mu(x)$ to interpolate the $\Sigma^*\bar{D}$ pentaquark molecular state; choose the component $\mathcal{S}^\alpha_{ud}\gamma_{\alpha}\gamma_{5}c\,\bar{c}\gamma_{\mu}u$ in the current  $J^4_\mu(x)$ to interpolate the $\Sigma\bar{D}^*$ pentaquark molecular state; choose the component $\mathcal{S}^{ud}_{\mu}c\,\bar{c}\gamma_{\nu}u$ in the current  $J^1_{\mu\nu}(x)$ to interpolate the $\Sigma^*\bar{D}^*$ pentaquark molecular state, see Eqs.\eqref{Fierz-J1}-\eqref{Fierz-Jmunu2}.

The experimental values of the masses of the LHCb pentaquark candidates $P_c(4312)$,  $P_c(4440)$ and  $P_c(4457)$ can be
reproduced  both in the diquark-diquark-antiquark type pentaquark scenario and in the baryon-meson molecule scenario. A possible interpretation is that the main Fock components of the $P_c(4312)$,  $P_c(4440)$ and  $P_c(4457)$ are the diquark-diquark-antiquark type pentaquark states, which  couple strongly to the baryon-meson pairs $\bar{D}\Sigma_c$, $\bar{D}\Sigma_c^*$, $\bar{D}^{*}\Sigma_c$  and   $\bar{D}^{*}\Sigma_c^*$, respectively, as the meson-baryon type currents chosen in Ref.\cite{WangPenta-IJMPA} already appear in the Fierz rearrangements of the pentaquark currents in Eqs.\eqref{Fierz-J1}-\eqref{Fierz-Jmunu2},  the strong couplings induce some pentaquark molecule  components, just-like in the mechanism  of the  $Y(4660)$.

In Ref.\cite{WangZG-Y4600-decay}, we  choose  the diquark-antidiquark  type tetraquark current interpolating  the $Y(4660)$  to study the  strong decays $Y(4660)\to J/\psi f_0(980)$, $ \eta_c \phi(1020)$,    $ \chi_{c0}\phi(1020)$, $ D_s \bar{D}_s$, $ D_s^* \bar{D}^*_s$, $ D_s \bar{D}^*_s$,  $ D_s^* \bar{D}_s$, $  \psi^\prime \pi^+\pi^-$, $J/\psi\phi(1020)$ with the QCD sum rules based on solid quark-hadron quality.
In calculations, we observe that the  hadronic coupling constants $ |G_{Y\psi^\prime f_0}|\gg |G_{Y J/\psi f_0}|$, which is consistent with the observation of the $Y(4660)$ in the $\psi^\prime\pi^+\pi^-$ mass spectrum, and favors the $\psi^{\prime}f_0(980)$ molecule assignment.

\section{Conclusion}
  In this article, we restudy the ground state mass spectrum of the diquark-diquark-antiquark type $uudc\bar{c}$ pentaquark states with the QCD sum rules by taking into account  all the
   vacuum condensates up to dimension $13$ in a consistent way  in  carrying out the operator product expansion, and use the energy scale formula $\mu=\sqrt{M_{P}-(2{\mathbb{M}}_c)^2}$ with the updated effective $c$-quark mass ${\mathbb{M}}_c=1.82\,\rm{GeV}$  to determine the ideal energy scales of the QCD spectral densities, and explore the possible assignments of the $P_c(4312)$, $P_c(4440)$ and $P_c(4457)$ in the scenario of the pentaquark states. The predicted masses  support assigning the $P_c(4312)$ to be the hidden-charm pentaquark state with $J^{P}={\frac{1}{2}}^-$, assigning the $P_c(4440)$ to be the hidden-charm pentaquark state with  $J^{P}={\frac{1}{2}}^-$, ${\frac{3}{2}}^-$ or ${\frac{5}{2}}^-$, assigning the $P_c(4457)$ to be the hidden-charm pentaquark state with  $J^{P}={\frac{1}{2}}^-$ or ${\frac{3}{2}}^-$.
  More experimental data and theoretical works are still needed to   identify the $P_c(4312)$, $P_c(4440)$ and $P_c(4457)$ unambiguously.

\section*{Acknowledgements}
This  work is supported by National Natural Science Foundation, Grant Number  11775079.


\begin{thebibliography}{99}

\bibitem{LHCb-4380} R. Aaij  et al, Phys. Rev. Lett. {\bf 115} (2015) 072001.

\bibitem{LHCb-Pc4312} R. Aaij et al, Phys. Rev. Lett. {\bf 122} (2019) 222001.


\bibitem{di-di-anti-penta-1} L. Maiani, A. D. Polosa and V. Riquer,  Phys. Lett. {\bf B749} (2015) 289.

\bibitem{di-di-anti-penta-2} V. V. Anisovich, M. A. Matveev, J. Nyiri, A. V. Sarantsev and A. N. Semenova, arXiv:1507.07652.

\bibitem{di-di-anti-penta-3} G. N. Li, M. He and X. G. He,  JHEP {\bf 1512} (2015) 128.

\bibitem{di-di-anti-penta-4} R. Ghosh, A. Bhattacharya and B. Chakrabarti, Phys. Part. Nucl. Lett. {\bf 14} (2017)  550.

\bibitem{di-di-anti-penta-5} V. V. Anisovich, M. A. Matveev, J. Nyiri, A. V. Sarantsev and A. N. Semenova, Int. J. Mod. Phys. {\bf A30} (2015) 1550190.


\bibitem{Wang1508-EPJC} Z. G. Wang, Eur. Phys. J. {\bf C76} (2016) 70.

\bibitem{WangHuang-EPJC-1508-12} Z. G. Wang  and T. Huang, Eur. Phys. J. {\bf C76} (2016)  43.

\bibitem{WangZG-EPJC-1509-12} Z. G. Wang, Eur. Phys. J. {\bf C76} (2016)  142.

\bibitem{WangZG-NPB-1512-32} Z. G. Wang, Nucl. Phys. {\bf B913} (2016) 163.

\bibitem{WangZhang-APPB} J. X. Zhang, Z. G. Wang and Z. Y. Di,  Acta Phys. Polon. {\bf B48} (2017) 2013.




\bibitem{Pc4312-penta-1} A. Ali and A. Y. Parkhomenko, Phys. Lett. {\bf B793} (2019) 365.

\bibitem{Pc4312-penta-2} H. Mutuk, Chin. Phys. {\bf C43} (2019)  093103.

\bibitem{Pc4312-penta-3} R. Zhu, X. Liu, H. Huang and C. F. Qiao,  Phys. Lett. {\bf B797} (2019) 134869.


\bibitem{di-tri-penta-1} R. F. Lebed, Phys. Rev. {\bf D92} (2015) 114030.

\bibitem{di-tri-penta-2} R. F. Lebed, Phys. Lett. {\bf B749} (2015) 454.

\bibitem{di-tri-penta-3} R. Zhu and C. F. Qiao, Phys. Lett. {\bf B756} (2016) 259.


\bibitem{mole-penta-1}  R. Chen, X. Liu, X. Q. Li and S. L. Zhu, Phys. Rev. Lett. {\bf 115} (2015) 132002.

\bibitem{mole-penta-2}  H. X. Chen, W. Chen, X. Liu, T. G. Steele and S. L. Zhu, Phys. Rev. Lett. {\bf 115} (2015)  172001.

\bibitem{mole-penta-3} L. Roca, J. Nieves and E. Oset,  Phys. Rev. {\bf D92} (2015)  094003.

\bibitem{mole-penta-4} J. He, Phys. Lett. {\bf B753} (2016) 547.

\bibitem{mole-penta-5} H. Huang, C. Deng, J. Ping and F. Wang, Eur. Phys. J. {\bf C76} (2016) 624.

\bibitem{mole-penta-6} F. K. Guo, U. G. Meissner, W. Wang  and Z. Yang, Phys. Rev. {\bf D92} (2015) 071502.

\bibitem{mole-penta-7} U. G. Meissner and J. A. Oller,   Phys. Lett. {\bf  B751} (2015) 59.

\bibitem{mole-penta-8}  T. J. Burns, Eur. Phys. J. {\bf A51} (2015)  152.

\bibitem{mole-penta-9}  K. Azizi, Y. Sarac and H. Sundu, Phys. Rev. {\bf D95} (2017)  094016.

\bibitem{mole-penta-10}  K. Azizi, Y. Sarac and H. Sundu, Phys. Lett. {\bf B782} (2018) 694.

\bibitem{WangPenta-IJMPA} Z. G. Wang,   Int. J. Mod. Phys. {\bf A34} (2019)  1950097.


\bibitem{Pc4312-mole-penta-1} R. Chen, Z. F. Sun, X. Liu and S. L. Zhu,  Phys. Rev. {\bf D100} (2019)  011502.

\bibitem{Pc4312-mole-penta-2} H. X. Chen, W. Chen and S. L. Zhu,  Phys. Rev. {\bf D100} (2019) 051501.

\bibitem{Pc4312-mole-penta-3} M. Z. Liu, Y. W. Pan, F. Z. Peng,  M. S. Sanchez, L. S. Geng, A. Hosaka and M. P. Valderrama, Phys. Rev. Lett. {\bf 122} (2019) 242001.

\bibitem{Pc4312-mole-penta-4} F. K. Guo, H. J. Jing, U. G. Meissner and S. Sakai, Phys. Rev. {\bf D99} (2019) 091501.

\bibitem{Pc4312-mole-penta-5} J. He, Eur. Phys. J. {\bf C79} (2019) 393.

\bibitem{Pc4312-mole-penta-6} C. J. Xiao, Y. Huang, Y. B. Dong, L. S. Geng and D. Y. Chen, Phys. Rev. {\bf D100} (2019)  014022.

\bibitem{Pc4312-mole-penta-7} Y. Shimizu, Y. Yamaguchi and M. Harada, arXiv:1904.00587.

\bibitem{Pc4312-mole-penta-8} H. Huang, J. He and J. Ping,  arXiv:1904.00221.

\bibitem{Pc4312-mole-penta-9} Z. H. Guo and J. A. Oller, Phys. Lett. {\bf B793} (2019) 144.

\bibitem{Pc4312-mole-penta-10} J. R. Zhang, arXiv:1904.10711.


\bibitem{Pc4312-hadrocharmonium} M. I. Eides, V. Y. Petrov and M. V. Polyakov, arXiv:1904.11616.


\bibitem{rescattering-penta-1}  M. Mikhasenko, arXiv:1507.06552.

\bibitem{rescattering-penta-2} X. H. Liu, Q. Wang and Q. Zhao,  Phys. Lett. {\bf B757} (2016) 231.

\bibitem{SVZ79}  M. A. Shifman, A. I. Vainshtein and V. I. Zakharov, Nucl. Phys. {\bf B147} (1979) 385, 448.

\bibitem{PRT85} L. J. Reinders, H. Rubinstein and S. Yazaki, Phys. Rept. {\bf 127} (1985) 1.

\bibitem{Nielsen-Review} R. M. Albuquerque, J. M. Dias, K. P. Khemchandani, A. Martinez Torres, F. S. Navarra, M. Nielsen and C. M. Zanetti,
J. Phys. {\bf G46} (2019)  093002.



\bibitem{WangEPJC-1601} Z. G. Wang, Eur. Phys. J. {\bf C76} (2016)  387.

\bibitem{Chung82} Y. Chung, H. G. Dosch, M. Kremer and D. Schall,  Nucl. Phys. {\bf B197} (1982) 55.

\bibitem{Bagan93} E. Bagan, M. Chabab, H. G. Dosch and S. Narison, Phys. Lett. {\bf B301}, 243 (1993).

\bibitem{Oka96} D. Jido, N. Kodama and M. Oka,  Phys. Rev. {\bf D54} (1996) 4532.

\bibitem{WangHbaryon-1} Z. G. Wang, Phys. Lett. {\bf B685} (2010) 59.

\bibitem{WangHbaryon-2} Z. G. Wang,  Eur. Phys. J. {\bf A45} (2010) 267.

\bibitem{WangHbaryon-3}    Z. G. Wang, Eur. Phys. J. {\bf C68} (2010) 459.

\bibitem{WangHbaryon-4}  Z. G. Wang, Eur. Phys. J. {\bf A47} (2011) 81.

\bibitem{WangHbaryon-5}    Z. G. Wang, Commun. Theor. Phys. {\bf 58} (2012) 723.



\bibitem{Wang-cc-baryon-penta} Z. G. Wang, Eur. Phys. J. {\bf C78} (2018) 826.

\bibitem{WangHuangTao} Z. G. Wang and T. Huang,  Phys. Rev. {\bf D89} (2014) 054019.


\bibitem{ColangeloReview} P. Colangelo and A. Khodjamirian, hep-ph/0010175.


\bibitem{PDG}  M. Tanabashi et al, Phys. Rev. {\bf  D98} (2018)  030001.


\bibitem{Narison-mix} S. Narison and R. Tarrach, Phys. Lett. {\bf 125 B} (1983) 217.



\bibitem{Wang-tetra-formula-1}  Z. G. Wang, Eur. Phys. J. {\bf C74} (2014)  2874.

\bibitem{Wang-tetra-formula-2}  Z. G. Wang and T. Huang, Nucl. Phys. {\bf A930} (2014) 63.


\bibitem{Wang-tetra-IJMPA-1} Z. G. Wang and Y. F. Tian, Int. J. Mod. Phys. {\bf A30} (2015) 1550004.

\bibitem{Wang-tetra-IJMPA-2} Z. G. Wang, Commun. Theor. Phys. {\bf 63} (2015) 325.

\bibitem{Wang-tetra-IJMPA-3} Z. G. Wang, Commun. Theor. Phys. {\bf 63} (2015) 466.


\bibitem{WangHuang-molecule-1} Z. G. Wang  and T. Huang, Eur. Phys. J. {\bf C74} (2014)  2891.

\bibitem{WangHuang-molecule-2} Z. G. Wang, Eur. Phys. J. {\bf C74} (2014)  2963.




\bibitem{WangLightDiquark} Z. G. Wang,  Commun. Theor. Phys. {\bf 59} (2013) 451.


\bibitem{WangDiquark-1} Z. G. Wang, Eur. Phys. J. {\bf C71} (2011) 1524.

\bibitem{WangDiquark-2}  R. T. Kleiv, T. G. Steele and A. Zhang, Phys. Rev. {\bf D87} (2013) 125018.


\bibitem{WangZG-Y4600-decay} Z. G. Wang, Eur. Phys. J. {\bf C79} (2019)  184.


\end{thebibliography}
\end{document}